\documentclass[useAMS,usenatbib]{mn2e}
\usepackage{longtable}
\usepackage{graphicx}
\usepackage{lscape}
\usepackage{rotating}
\newcommand{\mnras}{MNRAS}
\newcommand{\aj}{AJ}
\newcommand{\apj}{ApJ}
\newcommand{\apjs}{ApJ}
\newcommand{\apjl}{ApJ}
\newcommand{\aap}{A\&A}

\newcommand{\memsai}{Memorie della Societa Astronomica Italiana}

\newcommand{\Teff}{\mbox{$T_{\mathrm{eff}}$}}

\title[The  SDSS  WDMS binary  catalogue]  {White dwarf-main  sequence
  binaries from SDSS DR\,8: unveiling the cool white dwarf population}

\author[A.  Rebassa-Mansergas et al.]{A.  Rebassa-Mansergas$^{1,2}$,
  C. Agurto-Gangas$^2$, M. R.  Schreiber$^{2,3}$, B.T. G\"ansicke$^4$,
 \newauthor D. Koester$^5$\\
$^{1}$ Kavli Institute for Astronomy and Astrophysics, Peking University, Beijing 100871, China\\
$^{2}$ Departamento de F\'\i sica y Astronom\'\i a, Universidad de Valpara\'\i so, 
Avenida Gran Breta\~na 1111, Valpara\'\i so, Chile \\
$^{3}$ Millenium Nucleus "Protoplanetary Disks in ALMA Early Science,"
Universidad de Valpara\'\i so, Avenida Gran Breta\~na 1111, Valpara\'\i so, Chile\\
$^{4}$ Department of Physics, University of Warwick, Coventry CV4 7AL, UK \\
$^{5}$ Institut f\"ur Theoretische Physik und Astrophysik, University of Kiel,
24098 Kiel, Germany\\
}

\begin{document}
\date{Accepted 2013. Received 2013; in original form 2013}
\pagerange{\pageref{firstpage}--\pageref{lastpage}} \pubyear{2013}
\maketitle

\begin{abstract}
The  spectroscopic  catalogue  of  white  dwarf-main  sequence  (WDMS)
binaries from the  Sloan Digital Sky Survey (SDSS)  is the largest and
most  homogeneous  sample of  compact  binary  stars currently  known.
However, because of selection  effects, the current sample is strongly
biased against systems containing  cool white dwarfs and/or early type
companions,   which   are   predicted   to  dominate   the   intrinsic
population. In  this study we  present colour selection  criteria that
combines  optical ($ugriz$  DR\,8  SDSS) plus  infrared ($yjhk$  DR\,9
UKIRT Infrared  Sky Survey (UKIDSS),  $JHK$ Two Micron All  Sky Survey
(2MASS)  and/or $w_1w_2$ Wide-Field  Infrared Survey  Explorer (WISE))
magnitudes to  select 3\,419 photometric candidates  of harboring cool
white  dwarfs and/or  dominant (M  dwarf) companions.   We demonstrate
that 84  per cent of our  selected candidates are  very likely genuine
WDMS  binaries, and that  the white  dwarf effective  temperatures and
secondary star spectral  types of 71 per cent  of our selected sources
are expected  to be below  $\la$10\,000-\,15\,000\,K, and concentrated
at $\sim$M2-3,  respectively.  We also  present an updated  version of
the spectroscopic  SDSS WDMS  binary catalogue, which  incorporates 47
new systems  from SDSS DR\,8.  The  bulk of the  DR\,8 spectroscopy is
made up  of main-sequence stars and  red giants that  were targeted as
part of the Sloan Extension for Galactic Understanding and Exploration
(SEGUE)  Survey,  therefore  the  number  of  new  spectroscopic  WDMS
binaries  in  DR\,8 is  very  small  compared  to previous  SDSS  data
releases.  Despite their low number,  DR\,8 WDMS binaries are found to
be  dominated by systems  containing cool  white dwarfs  and therefore
represent  an important  addition  to the  spectroscopic sample.   The
updated SDSS  DR\,8 spectroscopic catalogue of  WDMS binaries consists
of  2316 systems.   We  compare our  updated  catalogue with  recently
published  lists  of WDMS  binaries  and  conclude  that it  currently
represents  the  largest,  most  homogeneous and  cleanest  sample  of
spectroscopic WDMS binaries from SDSS.
\end{abstract}


\label{firstpage}

\section{Introduction}
\label{s-intro}

White  dwarf-main sequence  binaries (WDMS)  are compact  binary stars
that  descend from  main sequence  binaries,  and are  among the  most
common compact  binary objects in  the Galaxy. The  majority ($\sim$75
per  cent) of  the  initial  main sequence  binaries  from which  WDMS
binaries  derive have orbital  separations wide  enough to  avoid mass
transfer interactions.   Therefore the primary (or  more massive) main
sequence  star evolves  as a  single star  and the  orbital separation
widens as a result of the  primary losing mass at the asymptotic giant
branch \citep{willems+kolb04-1}.  In the remaining ($\sim$25 per cent)
of  the  cases the  main  sequence stars  are  close  enough for  mass
transfer  to be  initiated via  Roche-lobe overflow  once  the primary
ascends  the red  giant branch  or  the the  asymptotic giant  branch.
Unstable  mass  transfer  to  the secondary  main  sequence  companion
generally   brings   the  system   into   a   common  envelope   phase
\citep[CE;][]{iben+livio93-1,  webbink07-1},   in  which  the  orbital
separation  dramatically decreases due  to drag  forces of  the binary
components  with the  material of  the envelope,  formed by  the outer
layers of the giant.  The energy  released due to the shrinkage of the
orbit   is   used   to   expel  the   envelope   \citep{davisetal10-1,
  zorotovicetal10-1,          ricker+taam12-1,          passyetal12-1,
  rebassa-mansergasetal12-2},   exposing  a   post-CE   binary  (PCEB)
composed of the  core of the giant, i.e.  the  future white dwarf, and
the   secondary   main  sequence   companion.    The  orbital   period
distribution  of WDMS binaries  is therefore  bimodal, with  the close
PCEBs   peaking  at   short   orbital  periods   of  $\sim$\,8   hours
\citep{miszalskietal09-1, nebotetal11-1} and  the systems that did not
evolve  through   a  CE  phase  at  much   wider  orbital  separations
\citep[orbital        periods       $>$100\,days,][]{willems+kolb04-1,
  farihietal10-1}.

After  the envelope  is  ejected,  PCEBs continue  to  evolve to  even
shorter  orbital  periods  through  angular  momentum  loss driven  by
magnetic braking  and/or gravitational wave  emission. Therefore PCEBs
may  either  undergo  a  second  phase of  CE  evolution  (leading  to
double-degenerate white  dwarfs), or enter a  semi-detached state (and
appear  as   cataclysmic  variables  or   super-soft  X-ray  sources).

Thanks to  the Sloan Digital  Sky Survey \citep[SDSS,][]{yorketal00-1,
  stoughtonetal02-1} the number of  WDMS binaries has increased from a
few   ten  \citep{schreiber+gaensicke03-1}   to   over  two   thousand
\citep{silvestrietal06-1, helleretal09-1, morganetal12-1, liuetal12-1,
  weietal13-1}.   The  latest version  of  our  WDMS binary  catalogue
\citep{rebassa-mansergasetal12-1}, based on  SDSS data release (DR) 7,
contains 2248 systems, and represents the most complete and homogeneous
sample.  Follow-up  observational studies  based on this  large sample
have led to the identification of  a large number of wide binaries and
close          PCEBs         \citep[e.g.][]{rebassa-mansergasetal07-1,
  schreiberetal08-1, schreiberetal10-1}  that are being  used to study
several different  and important aspects in  modern astrophysics (e.g.
providing  crucial constraints  on  current theories  of CE  evolution
\citep{davisetal10-1,   zorotovicetal10-1,  rebassa-mansergasetal12-2}
and     on     the     origin     of     low-mass     white     dwarfs
\citep{rebassa-mansergasetal11-1};   testing  theoretical  mass-radius
relations  of  both white  dwarfs  and  low-mass  main sequence  stars
\citep{nebotetal09-1, pyrzasetal09-1, pyrzasetal12-1, parsonsetal10-1,
  parsonsetal12-1,  parsonsetal12-3};  and  constraining  the  pairing
properties of main sequence stars \citep{ferrario12-1}).

The  currently known  population of  SDSS WDMS  binaries is  formed by
systems  observed  as part  of  the first  and  second  phases of  the
operation  of SDSS:  SDSS-I  and SDSS-II.   Whilst  SDSS-I focused  on
targeting   galaxies  and   quasars  \citep{adelman-mccarthyetal08-1},
SDSS-II carried out three different surveys \citep{abazajianetal09-1}:
the Sloan Legacy Survey that completed the original SDSS-I imaging and
spectroscopic goals;  the SEGUE  survey \citep[the SDSS  Extension for
  Galactic  Understanding   and  Exploration,][]{yannyetal09-1},  that
obtained additional  imaging over a large range  of Galactic latitudes
as  well  as spectroscopy  for  $\sim240\,000$  stars;  and the  Sloan
Supernova Survey  that carried  out repeat imaging  of the  300 square
degree southern  equatorial stripe to discover  and measure supernovae
and other variable objects.

Due to the overlap between the colours of WDMS binaries containing hot
white  dwarfs  and/or late-type  (M  dwarf)  companions  and those  of
quasars  \citep{smolcicetal04-1}, the  target  selection algorithm  of
both  SDSS-I and  the Legacy  Survey of  SDSS-II resulted  in  a large
number  of  WDMS binaries  with  available  SDSS spectroscopy.   SEGUE
additionally performed  a dedicated  survey for finding  WDMS binaries
containing cool  white dwarfs  and/or early-type M  dwarfs/late-type K
dwarfs  based  on colour  selection  criteria  developed  by our  team
\citep{schreiberetal07-1, rebassa-mansergasetal12-1}. However, despite
the success  of our SEGUE survey,  the 251 systems  identified in this
way represent only  $\sim$10 per cent of the  known population of SDSS
WDMS   binaries,  hence   remain   still  clearly   under-represented.
Moreover, it is important to bear  in mind that any WDMS binary sample
based on  optical colours/spectra alone,  such as the SDSS  sample, is
bound to be  incomplete as only binaries with  both components visible
at optical wavelengths can be identified.

The  aim of  this paper  is to  build on  the spectroscopic  SDSS WDMS
binary catalogue  by identifying WDMS binaries  within the photometric
footprint  of SDSS  DR\,8 without  the need  of SDSS  spectra,  and to
extend  the parameter  range of  the known  WDMS binary  population by
extending  the wavelength  range used  for their  identification, thus
overcoming the selection effects just described.  For this purpose, we
develop  colour selection  criteria  based on  a  combination of  SDSS
optical  plus   infrared  magnitudes  for   selecting  WDMS  binaries,
specifically focused on detecting systems containing cool white dwarfs
and/or  companions   dominating  the  system   luminosity,  which  are
predicted to  represent a large  fraction of the  intrinsic population
\citep{schreiber+gaensicke03-1}.   In  addition,  we  search  for  new
spectroscopic WDMS binaries observed by DR\,8.

\section{The photometric selection}
\label{s-phot}

We  select photometric  WDMS  binary candidates  following a  two-step
procedure.  First,  we apply colour  selection criteria based  on SDSS
$ugriz$ magnitudes,  which allows  us to exclude  single main-sequence
stars.  Second, we search  for available infrared excess detections of
our selected  candidates and apply  additional colour cuts based  on a
combination  of optical  plus infrared  magnitudes.   This efficiently
excludes  quasars from our  candidate list  and selects  WDMS binaries
dominated  by the  flux  of  the secondary  star,  i.e.  systems  that
contain cool  white dwarfs and/or  early-type (M dwarf)  main sequence
companions.

\subsection{Photometric selection of WDMS binaries in SDSS}
\label{s-sdssphot}

SDSS WDMS binaries form a ``bridge'' in colour space that connects the
white dwarf  locus to that of  low-mass stars \citep{smolcicetal04-1}.
Based  on this  bridge we  develop the  following colour  criteria for
selecting WDMS binaries within SDSS:

\begin{eqnarray}
   15 \leq g \leq 19, \\
(u - g)>-0.6, \\
-0.5<(g - r)<1.3, \\
 -0.4<(r - i)<1.6, \\
 -0.8<(i-z)<1.15
\end{eqnarray}

\begin{eqnarray}
(u - g)< 0.93-0.27\times(g - r)-4.7\times(g - r)^{2} \nonumber\\ 
+ 12.38\times(g - r)^{3}+3.08\times(g - r)^{4}-22.19\times(g - r)^{5}\nonumber \\
 +16.67\times(g - r)^6 - 3.89\times(g - r)^{7}
\end{eqnarray}

\begin{eqnarray}
   (g - r)<2\times(r - i)+0.38 \hspace{0.4cm}  \mathrm{if} \hspace{0.44cm} -0.4 <(r - i)\leq0.06 \\
   (g - r)<0.5 \hspace{0.6cm}  \mathrm{if} \hspace{0.4cm}  0.06<(r - i)\leq 0.3 \\
   (g - r)<4.5\times(r - i)-0.85 \hspace{0.6cm}  \mathrm{if} \hspace{0.4cm} 0.3<(r - i)\leq 0.48
\end{eqnarray}

\begin{eqnarray}
(r-i) < 0.5 + (i-z) \hspace{0.6cm}  \mathrm{if} \hspace{0.6cm} (i-z) \leq 0\\
(r-i) < 0.5 + 2\times(i-z) \hspace{0.6cm} \mathrm{if} \hspace{0.6cm} (i-z) > 0
\end{eqnarray}

\begin{figure}
\centering
\includegraphics[width=\columnwidth]{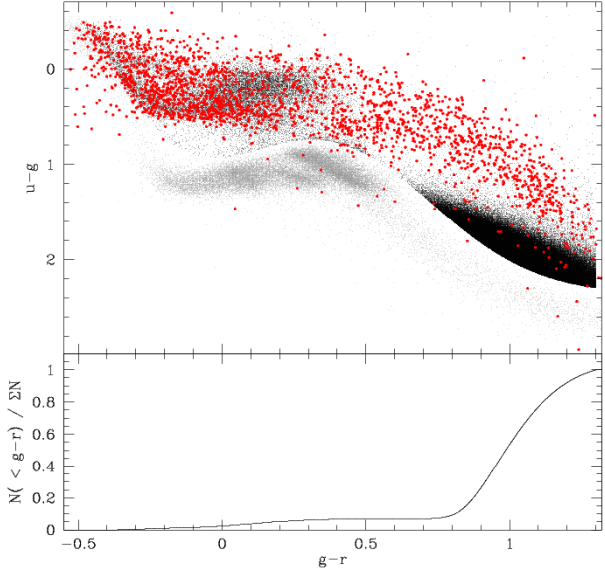}
\caption  {Top  panel:   photometrically  selected  SDSS  WDMS  binary
  candidates  in  the  $u-g$  vs.   $g-r$ plane  (black  solid  dots).
  Spectroscopically     confirmed    SDSS    WDMS     binaries    from
  \citet{rebassa-mansergasetal12-1}  are shown as  red solid  dots and
  main  sequence  stars  as  gray  solid dots.   Bottom  panel:  $g-r$
  cumulative distribution of the photometrically selected candidates.}
\label{f-first}
\end{figure}

We           use           the           $casjobs$           interface
\citep{li+thakar08-1} \footnote{http://skyserver.sdss3.sdss.org/CasJobs/}
to  select the  number  of  point sources  with  clean photometry  and
magnitude errors  below 0.1 within  the photometric data base  of SDSS
DR\,8,  satisfying  our  colour  selection.  This  search  results  in
953\,835 WDMS binary  candidates (see Table\,\ref{t-num}), illustrated
as  black solid  dots in  the  $u-g$ vs.   $g-r$ plane  (top panel  of
Figure\,\ref{f-first}).   For comparison, we  show also  main sequence
stars (gray solid dots)  and the 2248 spectroscopically confirmed WDMS
binaries      from      SDSS      DR\,7     (red      solid      dots,
\citealt{rebassa-mansergasetal12-1}).    In   the   bottom  panel   of
Figure\,\ref{f-first} we represent the $g-r$ cumulative distribution of
our 953\,835 selected sources.

A  close  inspection  of  the bottom  panel  of  Figure\,\ref{f-first}
reveals that  $g-r>$\,0.6 for  $\ga$\,90 per cent  of our  WDMS binary
candidates.   These systems  are concentrated  near the  main sequence
star locus (see top  panel of Figure\,\ref{f-first}), and overlap only
with  $\sim$3  per  cent   of  the  spectroscopically  confirmed  WDMS
binaries.  Inspection  of SDSS spectra of objects  falling within this
colour  space  reveals  $\sim$98  per  cent  of  these  systems  being
galaxies\footnote{It is intriguing that such a relatively large number
  of galaxies were flagged by  SDSS as point-sources, thus passing our
  initial  selection criteria.}.   We therefore  decide to  refine our
selection  criteria  to exclude  these  objects  from  our sample  and
re-write Equation\,6 as follows:

\begin{eqnarray}
(u - g)< 0.93-0.27\times(g - r)-4.7\times(g - r)^{2} \nonumber\\ 
+ 12.38\times(g - r)^{3}+3.08\times(g - r)^{4}\nonumber\\ 
-22.19\times(g - r)^{5} +16.67\times(g - r)^6 \nonumber \\
- 3.89\times(g - r)^{7} \hspace{0.6cm} \mathrm{if} \hspace{0.6cm} (g-r) \le 0.52,\\
(u-g) < 0.4 + (g-r) \hspace{0.6cm} \mathrm{if} \hspace{0.6cm} (g-r) > 0.52
\end{eqnarray}

\begin{figure*}
\centering
\includegraphics[angle=-90,width=0.3\textwidth]{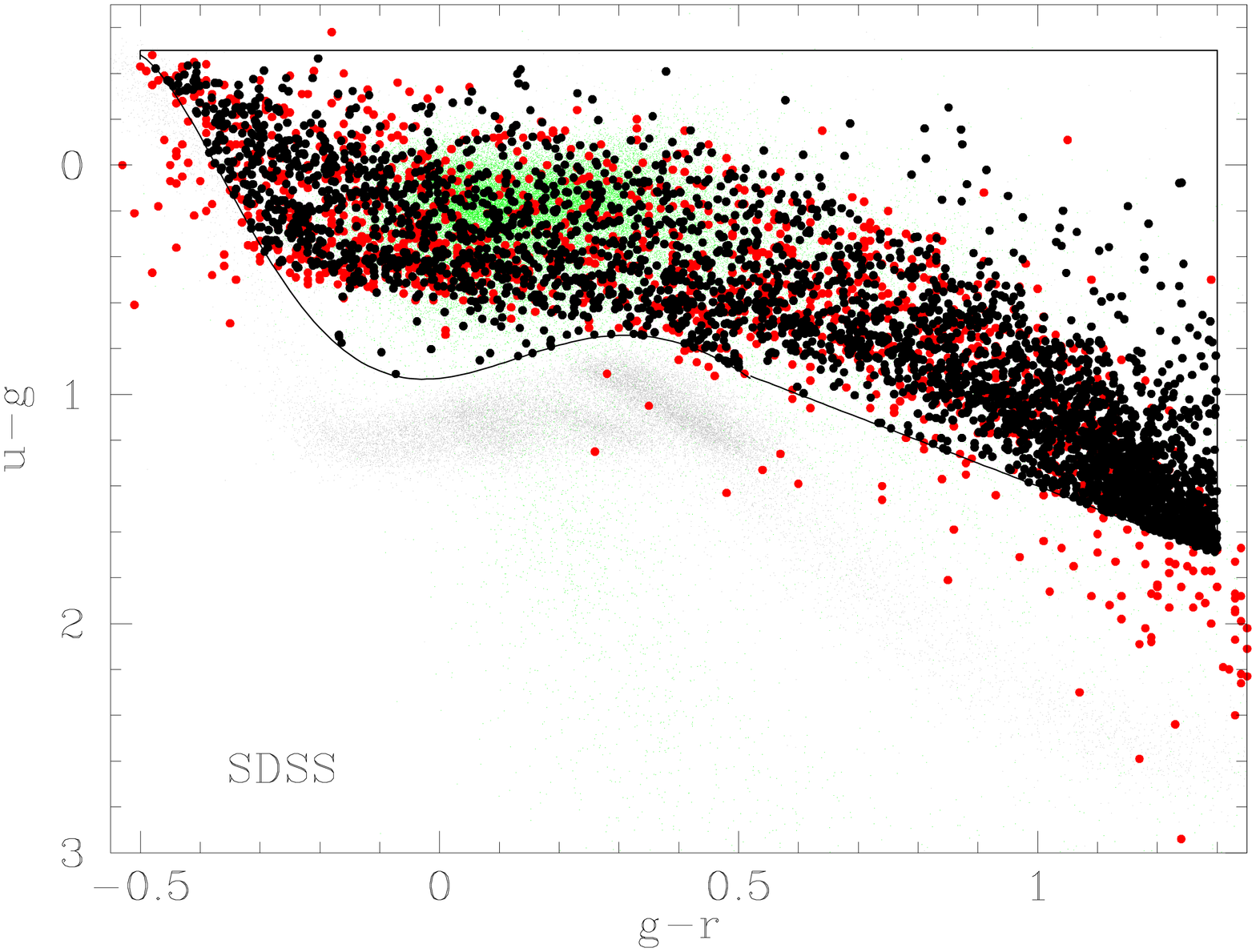}
\includegraphics[angle=-90,width=0.3\textwidth]{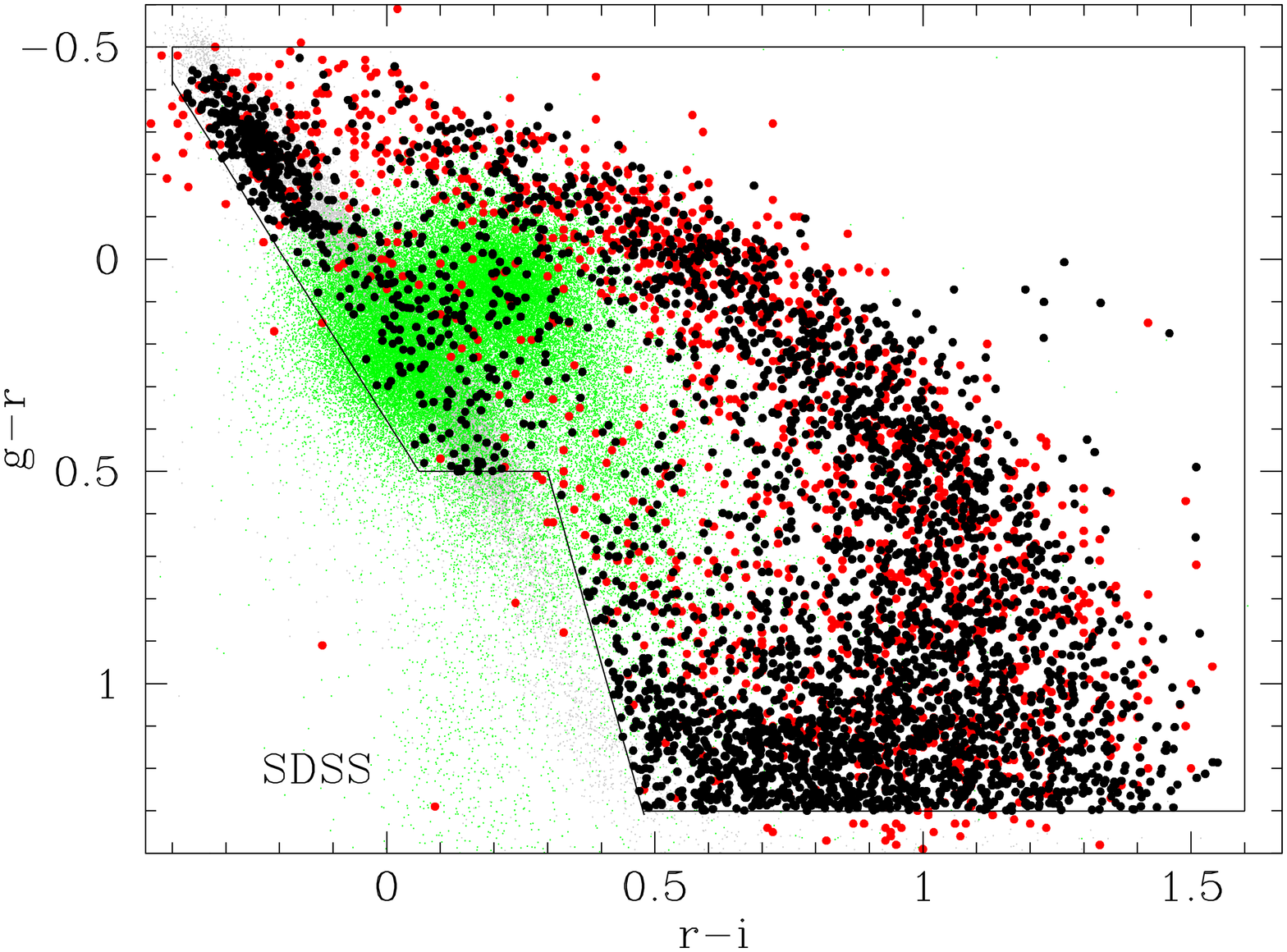}
\includegraphics[angle=-90,width=0.3\textwidth]{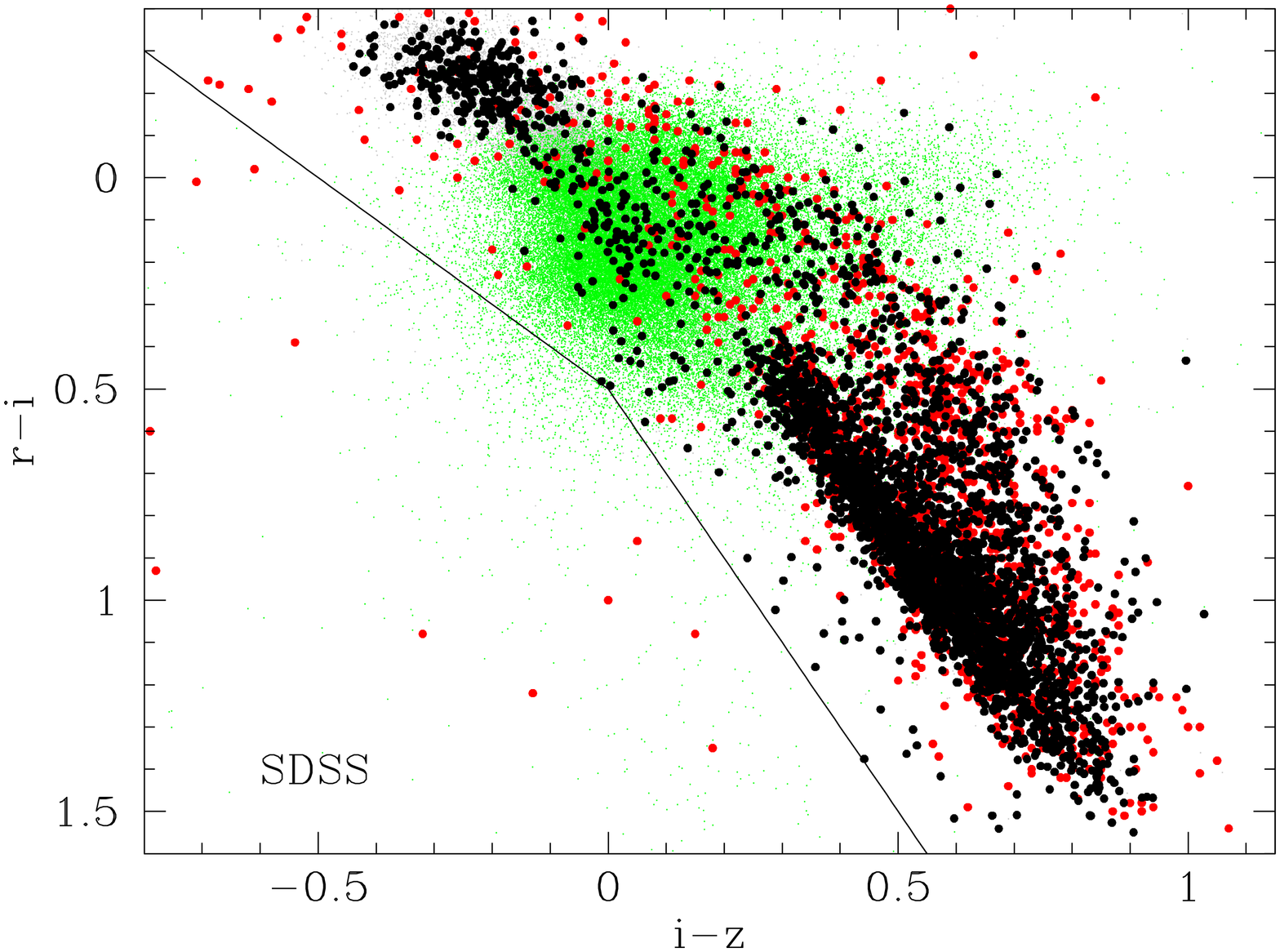}
\includegraphics[angle=-90,width=0.3\textwidth]{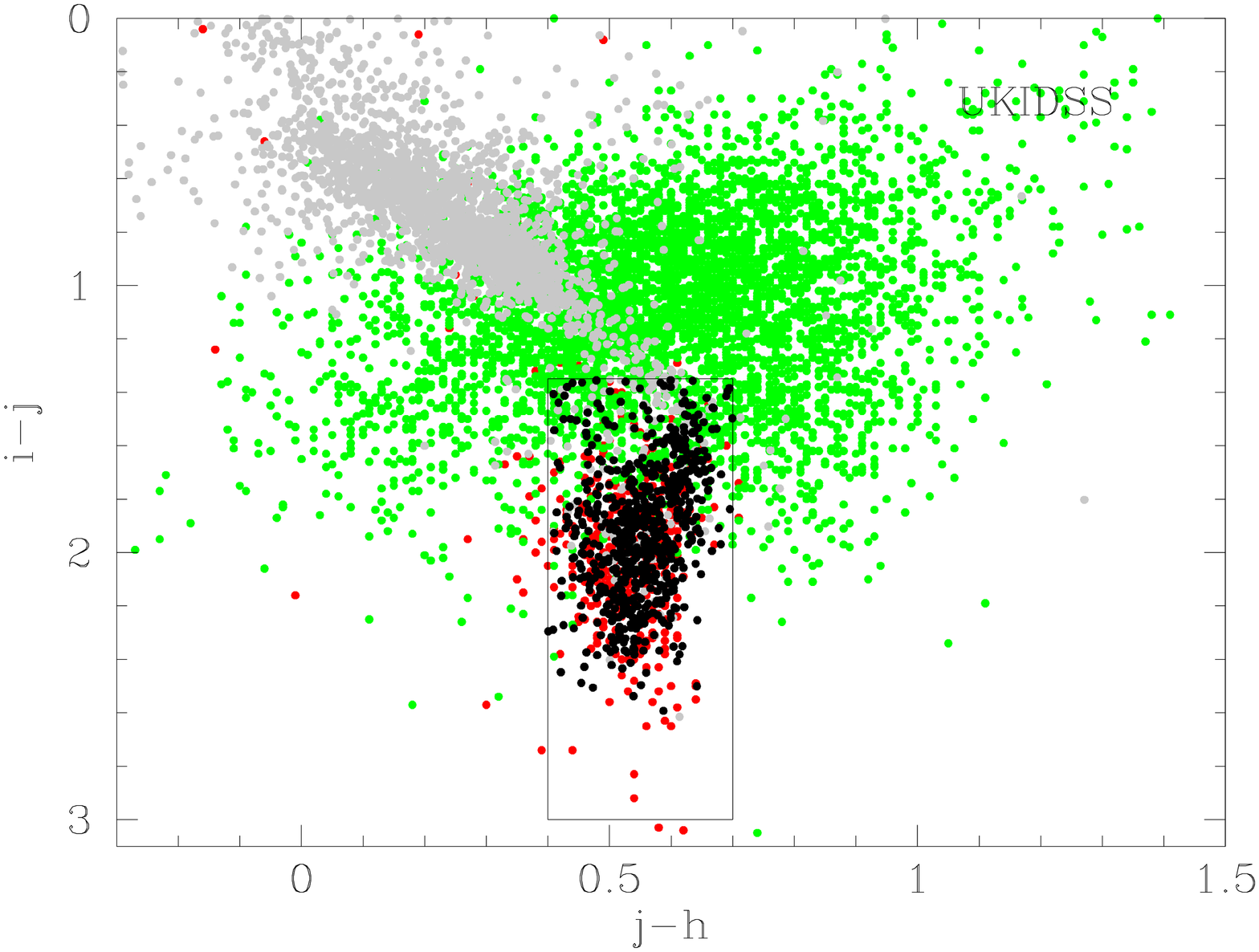}
\includegraphics[angle=-90,width=0.3\textwidth]{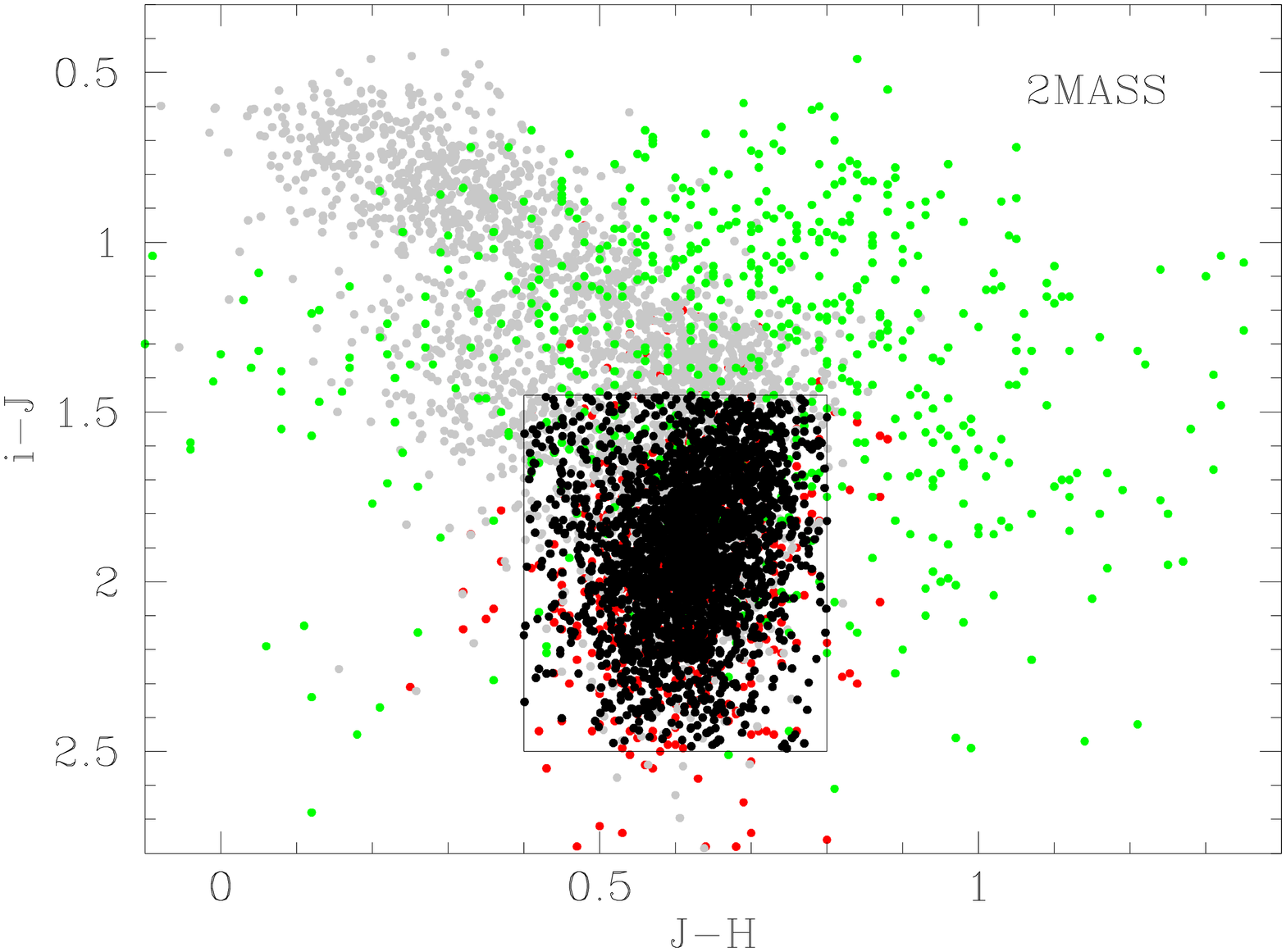}
\includegraphics[angle=-90,width=0.3\textwidth]{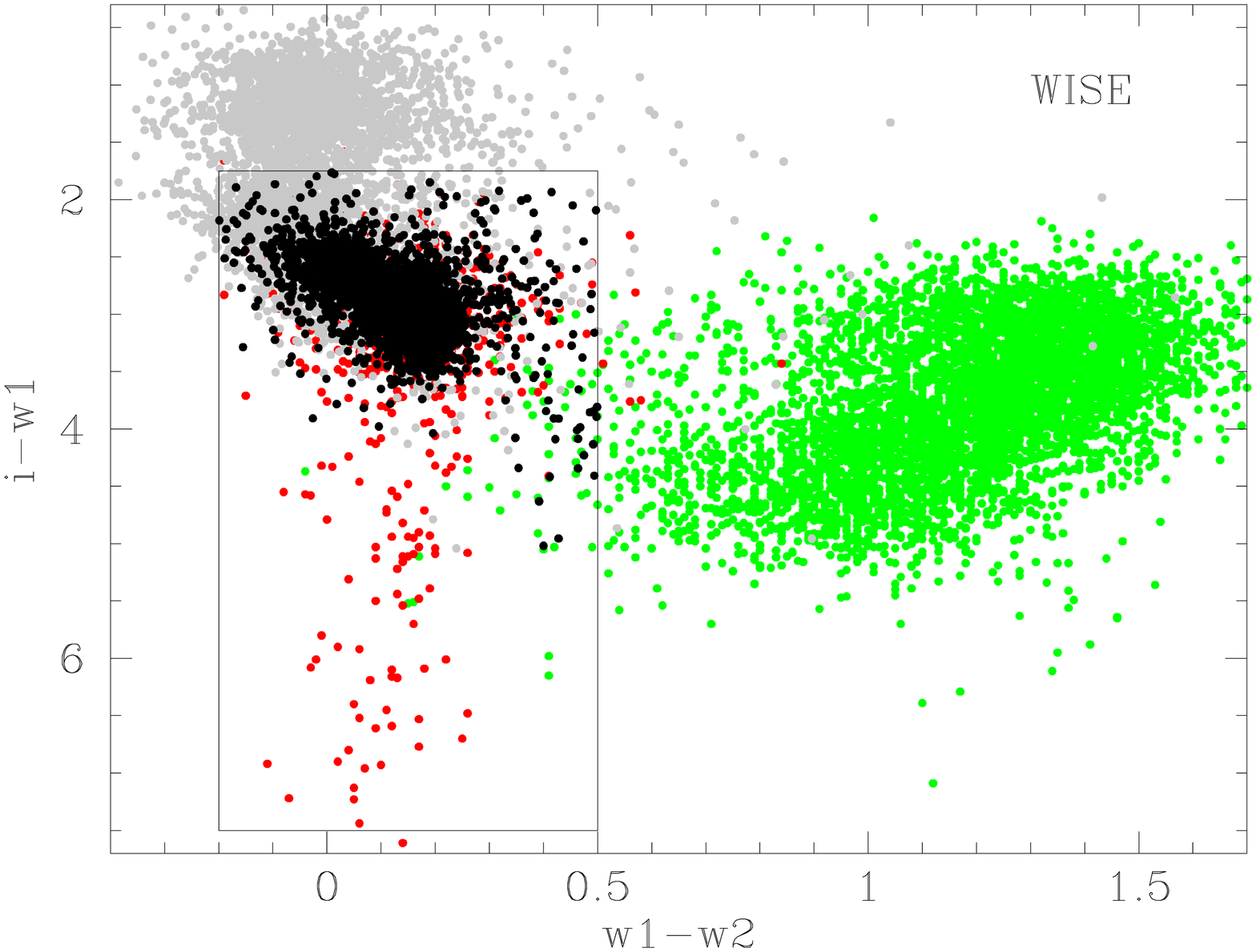}
\caption {Colour selection of WDMS binaries that combines optical SDSS
  and  infrared  UKIDSS, 2MASS  and/or  WISE  magnitudes (black  solid
  lines)  and the  3\,419 resulting  photometric WDMS  binary selected
  candidates  (black solid  dots).   To illustrate  how main  sequence
  stars  and  quasars  are  efficiently excluded  with  our  selection
  criteria    we    represent     a    sample    of    quasars    from
  \citet{schneideretal10-1}  as  solid green  dots  and main  sequence
  stars  as gray  solid dots.   Spectroscopically confirmed  SDSS WDMS
  binaries  from \citet{rebassa-mansergasetal12-1}  are also  shown as
  red  solid dots.}
\label{f-newcuts}
\end{figure*}

Our  refined  selection  criteria  are  shown on  the  top  panels  of
Figure\,\ref{f-newcuts} (black  solid lines) and reduce  the number of
photometrically  selected  WDMS  binary  candidates  to  67\,378  (see
Table\,\ref{t-num}).  For comparison,  main sequence stars (gray solid
dots), quasars (green solid dots) and spectroscopically confirmed SDSS
WDMS   binaries    (red   solid   dots)   are    also   shown.    From
Figure\,\ref{f-newcuts} (top  panels) it is apparent  that single main
sequence   stars   are    efficiently   excluded   with   our   colour
selection. However,  our sample is expected to  be highly contaminated
by  quasars.  Single  white dwarfs  are also  expected to  be  a large
source of contamination,  as white dwarf colours are  similar to those
of WDMS binaries in which  the white dwarf flux dominates the spectral
energy distribution.

\begin{table}
\centering
\caption{\label{t-num}  The number  of  photometrically selected  WDMS
  binary candidates  as function  of the progressively  refined colour
  selections. The colour cut in  each step is described in more detail
  in  the main  text.  The  final number  of  WDMS binary  photometric
  selected candidates is 3\,419.}  \setlength{\tabcolsep}{0.8ex}
\begin{small}
\begin{tabular}{ccc}
\hline
\hline
 N$_\mathrm{WDMS}$        & description          & excludes \\
\hline
953\,835  & SDSS colours         & main sequence stars  \\
67\,378   & refined SDSS colours & galaxies   \\
48\,163   & IR detections  &  objects with no IR          \\
4\,237    & optical plus infrared colours     & quasars    \\
3\,419    & comparison with WD catalogues   & white dwarfs \\
\hline
\end{tabular}
\end{small}
\end{table}

\subsection{Infrared excess}
\label{s-excess}

Here we search for available infrared excess detections of our 67\,378
selected  candidates   (Table\,\ref{t-num})  and  develop  additional
selection  criteria based on  a combination  of optical  plus infrared
magnitudes that allow  us to efficiently exclude quasars.   We do this
by   cross-correlating  our   list  with   the  $yjhk$   UKIDSS  DR\,9
\citep{lawrenceetal07-1,       warrenetal07-1},       $JHK$      2MASS
\citep{skrutskieetal06-1}  and  $w_1w_2$  WISE  \citep{wrightetal10-1}
surveys.   In all cases  we restrict  the search  to the  best quality
data, i.e.   we use quality flags AAA  in 2MASS and AA  in WISE.  This
results in 48\,163  objects with available detections in  at least one
of  the three  considered infrared  surveys  (Table\,\ref{t-num}).  We
further consider only objects satisfying the following conditions:

\begin{eqnarray}
1.35 < (i-j) < 3, \hspace{0.6cm} 0.4 < (j-h) < 0.7\\
1.45 < (i-J) < 2.5, \hspace{0.6cm} 0.4 < (J-H) < 0.8\\
1.75 < (i-w_1) < 7.5, \hspace{0.6cm} -0.2 < (w_1-w_2) < 0.5
\end{eqnarray}

The  above colour  criteria are  illustrated on  the bottom  panels of
Figure\,\ref{f-newcuts}  (black solid  lines) and  reduce the  list of
photometric  WDMS binary  candidates  to 4\,237  (Table\,\ref{t-num}).
Inspection  of Figure\,\ref{f-newcuts}  (bottom  panels) reveals  that
quasars are indeed efficiently excluded by our selection criteria.

\subsection{Comparison with white dwarf catalogues}

As mentioned  above, single  white dwarfs are  also expected to  be an
important  source  of   contamination  in  our  sample.   Fortunately,
comprehensive catalogues  of single white  dwarfs within the  SDSS DR7
footprint   are    available   \citep{girvenetal11-1,   debesetal11-1,
  kleinmanetal13-1}.  We  compare our WDMS binary  candidate list with
the catalogues  of \citet{kleinmanetal13-1} and \citet{girvenetal11-1}
and exclude all positive matches.  This reduces our number of selected
candidates to  3\,419 (Table\,\ref{t-num}), shown as  black solid dots
in  Figure\,\ref{f-newcuts}.  Coordinates,  optical SDSS  and infrared
UKIDSS,  2MASS  and  WISE  magnitudes of  the  3\,419  photometrically
selected      WDMS       binaries      can      be       found      in
Table\,\ref{t-photo}\footnote{Among  the  3\,419 selected  candidates,
  1\,109  have available  $yjhk$ UKIDSS  magnitudes,  2\,459 available
  $JHK$ 2MASS  magnitudes, 2\,606 available  $w_1w_2$ WISE magnitudes,
  and 501 available magnitudes from all three infrared surveys.}.

It is worth  mentioning that the photometric white  dwarf catalogue of
\citet{girvenetal11-1}  includes SDSS  white dwarfs  for  which UKIDSS
near-infrared excess is detected,  which implies we might be excluding
some  genuine WDMS binaries  in this  exercise.  However,  the overlap
between our WDMS  binary photometric candidate list and  the sample of
WDMS   binaries   that  may   be   included   in   the  catalogue   by
\citet{girvenetal11-1}  is  very small,  as  the  colour selection  by
\citet{girvenetal11-1} removes  any white dwarf that  has a noticeable
excess  in the SDSS  $i$ magnitude,  i.e.  the  main interest  of that
study is to  detect infrared excess typical of  debris discs around or
brown dwarf companions to white  dwarfs.  It is also important to keep
in mind that the single white dwarf catalogues are based on SDSS DR\,7
and that  we are  considering DR\,8 data  here.  Therefore, a  few per
cent of single  white dwarfs may still be included  in our WDMS binary
candidate list.

\begin{table*}
\centering
\caption{\label{t-photo}  Object names,  coordinates (in  degrees) and
  SDSS $ugriz$ , UKIDSS $yjhk$, 2MASS $JHK$ and WISE $w_{1}w_{2}$ magnitudes
  of the 3419 photometrically  selected WDMS binaries.  For those with
  available  spectroscopy,  we  include  the  spectral  classification
  (Table\,\ref{t-num1}).   The  complete  table  is available  in  the
  electronic edition of the paper.}  \setlength{\tabcolsep}{0.6ex}
\begin{small}
\begin{tabular}{cccccccccccccccccc}
\hline
\hline
SDSS\,J             &   ra    &   dec    &   $u$ &  $g$  &  $r$  &  $i$  &  $z$  &  $y$  &  $j$  &  $h$  &  $k$  &  $J$  &   $H$ &   $K$ &$w_{1}$ & $w_{2}$ &   type \\
\hline
000116.50+000204.8  & 0.31874 &  0.03468 & 18.85 & 18.80 & 18.91 & 19.04 & 19.22 & 18.90 & 18.94 &  0.00 &  0.00 &  0.00 &  0.00 &  0.00 &  0.00 &  0.00 &   unknown \\
000152.09+000644.5  & 0.46703 &  0.11236 & 19.03 & 18.57 & 17.89 & 17.48 & 17.17 & 16.51 & 16.05 & 15.40 & 15.28 &  0.00 &  0.00 &  0.00 &  0.00 &  0.00 &      WDMS \\
000218.67-064850.1  & 0.57781 & -6.81391 & 20.44 & 18.96 & 17.78 & 17.19 & 16.79 &  0.00 &  0.00 &  0.00 &  0.00 & 15.65 & 14.93 & 14.74 & 14.66 & 14.73 &         - \\
000238.16+162756.7  & 0.65898 & 16.46576 & 19.13 & 18.56 & 17.84 & 16.80 & 16.13 &  0.00 &  0.00 &  0.00 &  0.00 & 14.76 & 14.16 & 13.89 & 13.79 & 13.68 &         - \\
000356.94-050332.8  & 0.98723 & -5.05910 & 18.53 & 18.22 & 18.15 & 17.48 & 16.88 &  0.00 &  0.00 &  0.00 &  0.00 & 15.58 & 14.98 & 14.50 & 14.48 & 14.42 &      WDMS \\
000413.91+183616.4  & 1.05795 & 18.60455 & 18.71 & 18.53 & 18.57 & 17.87 & 17.23 &  0.00 &  0.00 &  0.00 &  0.00 & 15.84 & 15.16 & 14.96 & 14.79 & 14.61 &         - \\
000504.91+243409.6  & 1.27047 & 24.56934 & 19.51 & 18.89 & 18.47 & 17.48 & 16.81 &  0.00 &  0.00 &  0.00 &  0.00 & 15.36 & 14.72 & 14.40 & 14.26 & 14.11 &      WDMS \\
000541.94+133734.2  & 1.42474 & 13.62616 & 19.23 & 18.77 & 18.86 & 19.05 & 19.17 & 18.87 & 18.99 &  0.00 &  0.00 &  0.00 &  0.00 &  0.00 &  0.00 &  0.00 &         - \\
000553.76+113128.5  & 1.47402 & 11.52457 & 18.84 & 18.83 & 19.00 & 19.23 & 19.46 & 19.13 & 19.20 &  0.00 &  0.00 &  0.00 &  0.00 &  0.00 &  0.00 &  0.00 &         - \\
000559.88-054416.1  & 1.49948 & -5.73780 & 18.56 & 18.32 & 17.75 & 17.06 & 16.62 &  0.00 &  0.00 &  0.00 &  0.00 & 15.43 & 14.84 & 14.73 & 14.43 & 14.40 &      WDMS \\
\hline
\end{tabular}
\end{small}
\end{table*}

\subsection{The success rate of the photometric sample}

We have demonstrated that  our selection criteria are highly efficient
in  excluding single main  sequence stars  and quasars.   Single white
dwarfs  have  also  been  eliminated  from  our  list.   However,  the
remaining WDMS binary  candidate sample may contain a  small number of
other   astronomical   objects    with   similar   colours   such   as
e.g. cataclysmic variables.  It is therefore important to evaluate the
success  rate  in  identifying  WDMS binaries  among  our  photometric
candidates (hereafter success rate).  We do this as follows.

From our list of  3\,419 photometrically selected WDMS binary candidates
(Table\,\ref{t-num}),  567 have  SDSS spectroscopy  which  we visually
classify  as  follows  (Table\,\ref{t-num1}):  468 WDMS  binaries,  21
cataclysmic variables,  16 main  sequence stars, 2  main sequence-main
sequence  star  superpositions,  46  quasars, 2  main  sequence  stars
heavily affected  by extinction  and 12 objects  of unknown  type.  The
large number of WDMS binaries  in the spectroscopic sample indicates a
high   success   rate   ($\sim$83  per   cent,   Table\,\ref{t-num1}).
Apparently, a small  number of quasars that passed  our cuts, together
with some cataclysmic variables are the main sources of contamination.

\begin{figure}
\centering
\includegraphics[width=\columnwidth]{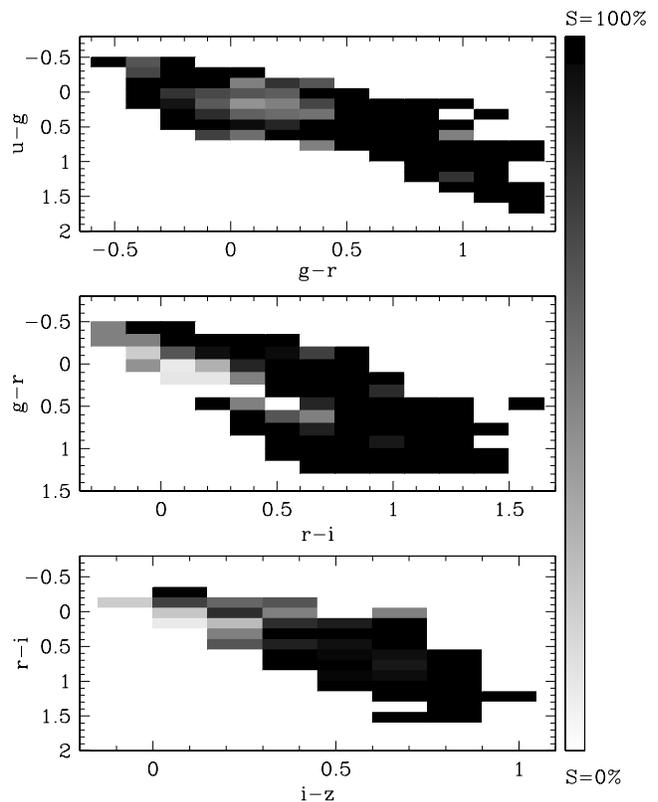}
\caption  {Density  maps  that  represent  the  success  rate  (S)  in
  selecting  WDMS  binaries  by  our  colour  criteria  in  the  $u-g$
  vs. $g-r$ (top panel), $g-r$  vs. $r-i$ (middle panel) and $r-i$ vs.
  $i-z$ (bottom panel) planes.}
\label{f-maps}
\end{figure}

It is to be expected that,  depending on the colour space, the success
rate varies, i.e. that the $\sim$83 per cent we obtained represents an
average  success  rate over  the  entire  colour  space. In  order  to
evaluate this  hypothesis we  divide the $u-g$  vs.  $g-r$,  $g-r$ vs.
$r-i$ and  $r-i$ vs.   $i-z$ planes into  cells of 0.15.   Within each
cell     we     calculate     the     local    success     rate     as
N$_\mathrm{wdms}$/N$_\mathrm{spec}$,   where   N$_\mathrm{wdms}$   and
N$_\mathrm{spec}$ are  the number  of spectroscopic WDMS  binaries and
the number  of systems with available  spectroscopy respectively.  The
dependence  of  the  success  rate  on location  in  colour  space  is
illustrated  in Figure\,\ref{f-maps}.   If we  take into  account only
cells with N$_\mathrm{spec}>20$, the  minimum and maximum success rate
we   obtain  are   43  per   cent  and   100  per   cent  respectively
(Figure\,\ref{f-maps}).   It  is also  evident  that  the success  rate
increases towards redder objects.

\begin{table}
\centering
\caption{\label{t-num1} Classification  of 567 WDMS  binary candidates
  in our list from  available SDSS spectroscopy.  The large percentage
  of genuine WDMS binaries indicates  a high success rate of our colour
  criteria in selecting WDMS binaries.}  \setlength{\tabcolsep}{0.8ex}
\begin{small}
\begin{tabular}{ccc}
\hline
\hline
 Type        & N          & percentage \\
\hline
WDMS       &  468   &     82.5\\
CV         &  21    &     3.7\\
MS         &  16    &     2.8\\
MS+MS sup. &  2     &     0.3\\
QSO        &  46    &     8.1\\
extinction MS &    2     &     0.3\\
unknown    &  12    &     2.4\\
\hline
\end{tabular}
\end{small}
\end{table}

\subsection{Characterization of the photometric sample}

\begin{figure}
\centering
\includegraphics[width=\columnwidth]{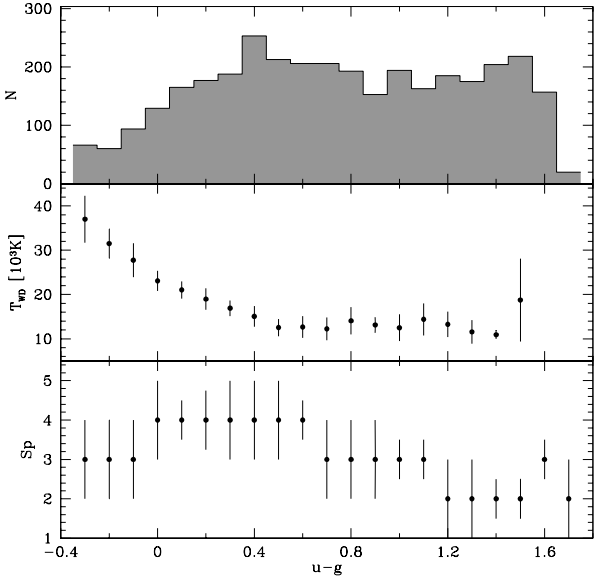}
\caption {Top panel: distribution  of the 3\,419 photometric SDSS WDMS
  binary   candidates   (Table\,\ref{t-num})    as   a   function   of
  $u-g$.  Middle  and bottom  panels:  average  white dwarf  effective
  temperatures  and secondary  star  (M dwarf)  spectral sub-types  of
  spectroscopically confirmed SDSS WDMS binaries also as a function of
  $u-g$.   The majority  of  our selected  photometric candidates  are
  expected  to be  WDMS binaries  with dominant  M dwarfs  and/or cool
  white dwarfs.}
\label{f-distr}
\end{figure}

Here  we investigate the  stellar properties  of our  photometric WDMS
binary candidate sample.  The top panel of Figure\,\ref{f-distr} shows
the distribution  of the 3\,419  selected sources (Table\,\ref{t-num})
as a function of $u-g$.  We calculate, for the same binning as for the
$u-g$ distribution,  the average (and standard  deviation) white dwarf
effective  temperatures  and  secondary  star spectral  types  of  the
spectroscopically     confirmed     SDSS     WDMS    binaries     from
\citet{rebassa-mansergasetal12-1},  and represent the  obtained values
on the middle and bottom panels of Figure\,\ref{f-distr} respectively.

The  white dwarf  effective temperatures  show a  clear  decrease with
increasing    $u-g$,   followed    by   a    flat    distribution   at
$\sim$10\,000-15\,000\,K for $u-g \ga  0.4$.  For values of $u-g>$1.3,
the spectroscopic  fits do not  provide reliable values for  the white
dwarf  effective  temperatures either  because  the  white dwarfs  are
extremely cool  and/or because the flux  of the M  dwarf dominates the
spectral energy distribution in  the SDSS spectra.  The spectral types
of the  secondary stars are of  $\sim$M3-4 for a broad  range of $u-g$
and show a trend towards  earlier spectral types ($\sim$M2) for $u-g >
1.2$.

This analysis  clearly demonstrates  that our colour  criteria selects
mainly ($\sim$ 70 per cent) WDMS binaries dominated by the flux of the
secondary star and/or containing cool white dwarfs ($\sim$M2-3, $\Teff
\la$10\,000-15\,000\,K), a population that is under-represented in the
current spectroscopic sample of SDSS WDMS binaries.

\begin{table*}
\centering
\caption{\label{t-num2} New spectroscopic SDSS WDMS binaries have been
  identified  in  this work  through  different  methods, outlined  in
  Section\,\ref{s-spec}  and summarised  in this  table.  In  the last
  column  we provide  the section  to refer  to for  further details.}
\setlength{\tabcolsep}{0.8ex}
\begin{small}
\begin{tabular}{ccccccccc}
\hline
\hline
  method      &  candidate WDMS  &  QSO  &  M dwarf &   WDMS &  DR\,8 WDMS  & total  WDMS  & total DR\,8 WDMS &\\
\hline
comparison with phot. sample           &      -    &  -   &   -      &   19      & 7   &  2267  & 7  & Section\,3.1\\
\hline                                                                
$\chi^2$-SN   fitting                  & 2353      & 18   & 456      &   44      &  37 &  2297  & 37 & Section\,3.2\\
M dwarf+GALEX; blue excess             & 456       &  -   &  -       &   8       &   8 &  2305  & 45 & Section\,3.3\\
\hline                                                                
completeness analysis                  & 2001      & 942  & 11       &   38      &  38 &  2306  & 46 & Section\,3.4\\
M dwarf+GALEX; blue excess             & 11        & -    &  -       &   0       &   0 &  2306  & 46 & \\
\hline
literature review                      &   -       & -    &  -       &   3       &   0 &  2309  & 46 & Section\,3.5\\
catalogue review                       &     -     & -    &  -       &   -2,+1   &   0 &  2308  & 46 & \\
\hline
comparison with \citet{liuetal12-1}    & 28        & -    &  -       &   0       &   0 &  2308  & 46 & Section\,3.5\\
comparison with \citet{morganetal12-1} & 85        & -    &  -       &   5       &   1 &  2313  & 47 & \\
comparison with \citet{weietal13-1}    & 281       & -    &  -       &   3       &   0 &  2316  & 47 & \\
\hline
\end{tabular}
\end{small}
   \end{table*}

\section{The spectroscopic catalogue}
\label{s-spec}

Our analysis  of the spectroscopic DR7 resulted  in the identification
of 2248 WDMS binaries \citep{rebassa-mansergasetal12-1}. Here, we test
both if  our photometric selection  (Section\,\ref{s-phot}) identifies
WDMS binaries  that we missed  in our spectroscopic  classification of
DR\,7,  and  if there  are  any  new  spectroscopic WDMS  binaries  in
DR\,8.  We first cross-correlate  the 468  systems in  our photometric
sample that have available SDSS spectra (Table\,\ref{t-num1}) with the
2248 confirmed WDMS binaries  from DR\,7, and incorporate all possible
missing objects to  the list.  We then follow  the methods outlined by
\citet{rebassa-mansergasetal10-1}    to    search    for    additional
spectroscopic WDMS binaries within DR\,8.

\subsection{Comparison with the photometric sample}
\label{s-cross}

19  WDMS binaries  with available  SDSS spectra  from  our photometric
sample  (Section\,\ref{s-phot})  are  not  included  in  the  list  by
\citet{rebassa-mansergasetal12-1}.  Of these,  seven are part of DR\,8
and    are    therefore    new    additions    to    the    catalogue:
SDSSJ082845.07+133551.0,                       SDSSJ093011.64+095319.5,
SDSSJ101356.32+272410.6,                       SDSSJ102122.45+433633.1,
SDSSJ102627.48+384502.4,                       SDSSJ121150.94+110543.2,
SDSSJ145722.85-012121.2).  

The remaining  12 systems are from  DR\,7 and are divided  into:
\begin{itemize}
\item five  M dwarf dominated  WDMS binaries (SDSSJ014113.10-084831.0,
  SDSSJ083833.17+140332.1,                     SDSSJ151251.47+010201.2,
  SDSSJ212125.32+010541.6, SDSSJ235143.39+362736.6).
\item six apparently single M  dwarfs with strong Balmer emission lines
  due to a (likely) close white  dwarf that heats the surface of the M
  dwarf  and/or  due   to  magnetic  activity  \citep{tappertetal11-1,
    rebassa-mansergasetal13-1}                (SDSSJ054251.34+010206.8,
  SDSSJ074645.01+425327.4,                     SDSSJ080239.07+102026.0,
  SDSSJ090210.97+252913.5,                     SDSSJ093127.22+151855.0,
  SDSSJ102804.59+081321.9).
\item one G star plus (hot) white dwarf (SDSSJ142838.99+424024.8).
\end{itemize}

In all  12 cases  the SDSS  spectra do not  reveal strong  white dwarf
features,   which  explains   why   these  systems   were  missed   by
\citet{rebassa-mansergasetal12-1}. The  addition of the  19 systems to
the  spectroscopic SDSS  WDMS binary  catalogue raises  the  number of
spectroscopic SDSS WDMS binaries to 2267 (Table\,\ref{t-num2}).

\subsection{The template-fitting method}
\label{s-tempfit}

Here we  identify new DR\,8  SDSS WDMS binaries following  the routine
developed  by  \citet{rebassa-mansergasetal10-1}.   This technique  is
based on reduced $\chi^{2}$ template fitting all new DR\,8 spectra, as
well as on signal-to-noise  (S/N) ratio constraints.  The template set
consists  of 163 spectra  of previously  confirmed SDSS  WDMS binaries
covering  a broad  range  of white  dwarf  effective temperatures  and
spectral subtypes (DA,  DB and DC) as well  as companion star spectral
types (M0-M9).  In practice, an equation of the form

\begin{equation}
\chi^2_\mathrm{max} = a \times (\mathrm{S/N_\mathrm{spec}})^b
\end{equation}
\noindent
is defined for each  of the 163 templates, where $\chi^2_\mathrm{max}$
is  the  maximum  $\chi^2$   allowed  between  the  template  and  the
considered SDSS  spectrum, S/N$_\mathrm{spec}$ is  the signal-to-noise
ratio of the SDSS spectrum, and $a$ and $b$ are fixed values that vary
from  template  to template.   All  objects  falling  below the  curve
defined for  each template are considered WDMS  binary candidates.  In
the process of WDMS binary  search we inspect the available SDSS DR\,8
images  of the selected  candidates in  order to  detect morphological
problems.  This can  be the case when single white  dwarfs or M dwarfs
are located close  to very bright stars that  cause scattered light to
enter the spectroscopic fibre  and result in an apparent two-component
spectrum; other cases can be the superposition along the line of sight
between two stars.   All these objects are removed  from our candidate
list.

We  obtain  a  list  of  2\,353  WDMS  binary  candidates  and  visual
inspection of  the spectra of  these candidates results in  44 genuine
WDMS  binaries (Table\,\ref{t-num2}).  Of  the 44  identified systems,
seven  objects are  included in  the list  of DR\,7  WDMS  binaries by
\citet{rebassa-mansergasetal12-1}, therefore only 37 are new additions
(these    include   the    seven   WDMS    binaries    identified   in
Section\,\ref{s-cross}).  The majority  of the rejected candidates are
early type main sequence stars  plus a noticeable fraction of M dwarfs
(456, see  Table\,\ref{t-num2}).  We  identify only 18  quasar spectra
and no spectra with white dwarf  features.  The sample of 456 M dwarfs
is further  investigated in the next  section, as a  fraction of these
might be WDMS binaries containing cool/unseen white dwarfs.

The relatively  small number of WDMS binaries  with spectroscopy found
within  DR\,8,  compared  to   earlier  data  releases,  is  a  direct
consequence of the spectroscopic  DR\,8 being dominated by single main
sequence  stars  and  giants  that  were observed  as  part  of  SEGUE
\citep{aiharaetal11-1, eisensteinetal11-1}.

\subsection{Identification of blue excess}

Searching  for WDMS binaries  within the  spectroscopic DR\,8  we have
identified  456  apparently  single  M  dwarfs  (Table\,\ref{t-num2}).
These may  be genuine M dwarfs  or WDMS binaries  containing very cool
and/or  unseen  white  dwarfs.   We  here follow  two  procedures  for
identifying white  dwarf primaries that contribute very  little to the
optical flux of the binaries.

\begin{figure}
\includegraphics[angle=-90,width=\columnwidth]{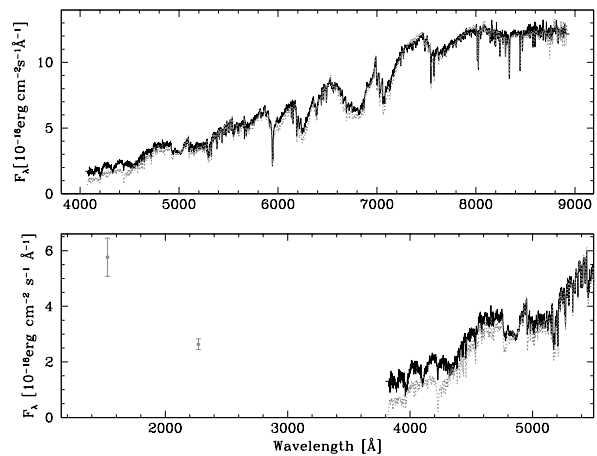}
\caption{\label{f-chisn1} SDSS  spectrum of SDSSJ\,011004.70+071702.8,
  an M dwarf  dominated WDMS binary. The presence of  a white dwarf is
  confirmed by template fitting the  spectrum with an M2 template (top
  panel, gray dotted line) and  by the GALEX near- and far-ultraviolet
  detections (bottom panel, gray solid dots).}
\end{figure}

First, we  cross-correlate our  list of 456  M dwarf  selected spectra
with the  near- and far-ultraviolet magnitudes provided  by the Galaxy
Evolution      Explorer     \citep[GALEX     DR\,6,][]{martinetal05-1,
  morrisseyetal05-1, seibertetal12-1}.  A clear  excess in the near or
far-ultraviolet GALEX  fluxes confirms the  presence of a  white dwarf
primary that  contributes too little  to be unambiguously  detected at
optical  wavelengths.  An  example is  shown  on the  bottom panel  of
Figure\,\ref{f-chisn1}.

An  additional  way of  identifying  cool  and/or  unseen white  dwarf
primaries is  by searching  for blue excess  in the M  dwarf dominated
spectra. We do this by fitting the spectra with the M0 to M9 templates
of   \citet{rebassa-mansergasetal07-1}.   We  calculate   the  reduced
$\chi^{2}$ between the M dwarf  spectra and the best-fit template over
the        4000-5000\,\AA\,($\chi^2_\mathrm{b}$)        and        the
7000-9000\,\AA\,($\chi^2_\mathrm{r}$) wavelength ranges.  Objects with
$\chi^2_\mathrm{b}/\chi^2_\mathrm{r}>2$ are  selected as WDMS binaries
containing    unseen   white   dwarfs.     However,   as    shown   by
\citet{rebassa-mansergasetal10-1},  this method  is likely  to provide
false detections in case of  low signal-to-noise ratio spectra, and is
prone to  select active  M dwarfs. We  therefore visually  inspect the
spectra of  all blue excess candidates.   An example of  a WDMS binary
recovered   in   this   way   is   shown   on   the   top   panel   of
Figure\,\ref{f-chisn1}.

The  search for  blue excess  and positive  detections in  GALEX DR\,6
confirms the presence of white  dwarf primaries in eight systems among
the 456  single M dwarf  candidate spectra, thus increasing  the total
number of spectroscopic WDMS binaries to 2\,305 (Table\,\ref{t-num2}).

\subsection{Catalogue completeness}
\label{s-internal}

We  evaluate here  the internal  completeness  of the  new SDSS  DR\,8
spectroscopic sample of WDMS binaries,  i.e.  the fraction of all WDMS
binaries contained in  the DR\,8 spectroscopic data base  that we have
identified.             In           \citet{rebassa-mansergasetal10-1,
  rebassa-mansergasetal12-1} we approached this aim by analysing small
areas in colour space that  are representative of the SDSS WDMS binary
population. Here  we base our study  on the entire  WDMS binary bridge
\citep{smolcicetal04-1}.  This is motivated  by the fact that the bulk
of the DR\,8 spectroscopy is made up of single main sequence and giant
stars obtained as  part of SEGUE that can  be easily excluded applying
the colour selection criteria defined by Equations\,1-11.

We use  the casjobs  interface to detect  the number of  point sources
with clean  photometry and available  DR\,8 SDSS spectra  that satisfy
Equations\,1-11.  Note  that in this  exercise we restrict  the search
for spectroscopic sources  that form part \emph{only} of  the new data
released  by  DR\,8.   This  search  results  in  2\,001  WDMS  binary
candidates  (Table\,\ref{t-num2}).  Visual  inspection  of the  2\,001
candidate  spectra reveals  38 WDMS  binaries,  942 quasars  and 11  M
dwarfs that might be genuine  M stars or WDMS binaries containing cool
and/or unseen white dwarf primaries (Table\,\ref{t-num2}, the majority
of  the  remaining  spectra   are  main  sequence  stars  affected  by
extinction, main  sequence-main sequence star  superpositions, and some
early type main  sequence stars).  Among the 11  M dwarf spectra, none
are  confirmed as WDMS  binaries due  to blue  excess detected  in the
spectrum or as resulting from positive GALEX DR\,6 detections.

All   WDMS   binaries  identified   in   this   exercise  except   one
(SDSSJ155847.38+431308.4,  an  M  dwarf  dominated  WDMS  binary)  are
included in our DR\,8 WDMS  binary list, thus providing a completeness
of $\sim$98 per cent.  We add SDSSJ155847.38+431308.4 to our catalogue
and the number of  spectroscopically confirmed SDSS WDMS binaries thus
raises to 2306.

\begin{table*}
\centering
\caption{\label{t-plt}  SDSS  plate, modified  Julian  date (MJD)  and
  fibre  identifiers of  45 DR\,7  WDMS  binary spectra  that are  not
  included    in    the     DR\,7    WDMS    binary    catalogue    by
  \citet{rebassa-mansergasetal12-1}.  These  spectra were taken during
  observations  performed  for  DR\,7,  however,  made  only  publicly
  available within DR\,8.}  \setlength{\tabcolsep}{0.8ex}
\begin{small}
\begin{tabular}{cccccccccccc}
\hline
\hline
 Object                   & plt  & MJD   & fib  & Object                    & plt  & MJD   & fib  & Object                   & plt  & MJD   & fib  \\
\hline
 SDSSJ002157.91-110331.6  & 1913 & 53295 &  242 &  SDSSJ102857.79+093129.9  & 2854 & 54465 &  576 &  SDSSJ155808.50+264225.8  & 2474 & 54333 &  598  \\
 SDSSJ011226.93+251149.8  & 2060 & 53388 &  274 &  SDSSJ104219.08+442916.0  & 2567 & 54179 &  636 &  SDSSJ161239.06+455132.3  &  814 &  5235 &  403  \\
 SDSSJ014143.68-093811.7  & 2865 & 54503 &  170 &  SDSSJ105346.29+291652.6  & 2359 & 53800 &  413 &  SDSSJ165112.44+415139.2  &  631 & 52054 &  474  \\
 SDSSJ030904.82-010100.9  &  412 & 51871 &  204 &  SDSSJ105421.97+512254.2  &  876 &  5234 &  533 &  SDSSJ171301.85+625135.9  &  352 & 51694 &  224  \\
 SDSSJ081716.98+054223.7  & 1296 & 52738 &  307 &  SDSSJ105617.52+505321.2  &  876 &  5234 &  590 &  SDSSJ171955.23+625106.8  &  352 & 51694 &  163  \\
 SDSSJ081959.21+060424.2  & 1296 & 52738 &  196 &  SDSSJ110750.15+050559.0  &  581 & 52353 &  495 &  SDSSJ172433.70+623410.0  &  352 & 51694 &  59   \\
 SDSSJ083410.38+135355.8  & 2427 & 53800 &  314 &  SDSSJ114312.57+000926.5  &  283 & 51660 &  437 &  SDSSJ172831.84+620426.5  &  352 & 51694 &  7    \\
 SDSSJ090455.46+184741.9  & 2285 & 53687 &  166 &  SDSSJ114312.57+000926.5  &  283 & 51584 &  433 &  SDSSJ173101.49+623316.0  &  352 & 51694 &  16   \\
 SDSSJ092737.67+255423.0  & 2294 & 54524 &  353 &  SDSSJ131334.74+023750.8  &  525 & 52029 &  509 &  SDSSJ173727.27+540352.2  &  360 & 51780 &  165  \\
 SDSSJ100529.93+521937.9  &  903 &  5238 &  169 &  SDSSJ132040.28+661214.8  &  496 & 51973 &  195 &  SDSSJ174214.72+541845.1  &  360 & 51780 &  115  \\
 SDSSJ100609.18+004417.1  &  269 & 51581 &  605 &  SDSSJ135930.96-101029.7  & 2716 & 54628 &  211 &  SDSSJ204117.50-062847.1  &  634 & 52149 &  47   \\
 SDSSJ101614.70+490930.4  &  873 &  5234 &  399 &  SDSSJ140723.04+003841.7  &  302 & 51616 &  464 &  SDSSJ222108.46+002927.7  & 1143 & 52592 &  462  \\
 SDSSJ101722.72+025147.8  &  574 & 52347 &  44  &  SDSSJ141220.70+654123.3  &  498 & 51973 &  545 &  SDSSJ222822.74+391239.8  & 2620 & 54339 &  73   \\
 SDSSJ101722.72+025147.8  &  574 & 52356 &  44  &  SDSSJ143746.70+573706.1  &  790 &  5234 &  120 &  SDSSJ224038.38-093541.4  &  722 & 52206 &  150  \\
 SDSSJ101722.72+025147.8  &  574 & 52366 &  50  &  SDSSJ150231.66+011046.0  &  310 & 51616 &  423 &  SDSSJ225334.79-090554.0  &  724 &  5223 &  433  \\
\hline
\end{tabular}
\end{small}
\end{table*}

\section{The final spectroscopic SDSS DR8 WDMS binary catalogue}
\label{s-final}

In  this work we  have raised  the number  of spectroscopic  SDSS WDMS
binaries to  2306 (Table\,\ref{t-num2}).  To this list  we include the
recently  discovered WDMS  binaries  SDSSJ013851.54-001621.6 \citep[][
  which   contains   an   ultra-cool   white   dwarf]{parsonsetal12-3},
SDSSJ135523.92+085645.4 \citep[][which contains  a hot white dwarf and
  a    likely     brown    dwarf    companion]{badenesetal13-1}    and
SDSSJ013532.97+144555.9  \citep[][   which  contains  a   brown  dwarf
  companion]{steeleetal13-1}.  In these  three cases, the SDSS spectra
are totally dominated by one of the stellar components, making it very
difficult for  our template fitting routine to  identify these systems
as WDMS binaries.  In  addition, we include SDSSJ053317.31-004321.9, a
resolved K  star plus  white dwarf binary  that we  previously missed.
Finally,  we exclude from  our catalogue  SDSSJ014349.22+002130.0, for
which the blue excess detected by \citet{rebassa-mansergasetal10-1} is
likely to  come from a  nearby quasar rather  than a white  dwarf, and
SDSSJ144335.19+004005.9, which is an eclipsing binary containing two M
dwarfs with an orbital  period of $\sim$1\,day (S.G.  Parsons, private
communication).  This  brings the  total number of  spectroscopic SDSS
WDMS binaries to 2308 (Table\,\ref{t-num2}).

In the course  of writing this paper three sets  of SDSS WDMS binaries
have  been  published by  three  different groups  \citep{liuetal12-1,
  morganetal12-1, weietal13-1}, and we compare our set of 2308 systems
with these three samples in what follows.

\subsection{Comparison with \citet{liuetal12-1}}

We detect all 523 WDMS binaries from \citet{liuetal12-1} except for 28
systems  that  we  do  not  consider  WDMS  binaries  after  a  visual
inspection of  the SDSS spectra.  An updated  classification for these
systems is given in Table\,A1.

\subsection{Comparison with \citet{morganetal12-1}}

A comparison  with the 1756 systems  of \citet{morganetal12-1} reveals
that  we are  missing 85  objects.   Inspecting the  SDSS spectra  and
images  of  these 85  systems  we  classify  80 as  non-WDMS  binaries
(Table\,A2).   The remaining  five systems  are:

\begin{itemize}
\item SDSSJ000421.61+004341.5, a DR\,7 noisy spectrum of a M star plus
  some blue excess.
\item SDSSJ083833.17+140332.1,  an M dwarf dominated  WDMS binary from
  DR\,8.
\item SDSSJ100413.18+342950.8, a white dwarf plus (likely) late K star
  companion from DR\,7.
\item  SDSSJ153648.31+010249.1, a  white dwarf  dominated  WDMS binary
  from DR\,7.
\item SDSSJ220436.50-002313.7, a noisy  spectrum of a WDMS binary from
  DR\,7.
\end{itemize}

We add these five systems to  our catalogue list.  The total number of
spectroscopic SDSS WDMS binaries thus raises to 2313 and the number of
new DR\,8 WDMS binaries increases to 47 (Table\,\ref{t-num2}).

\subsection{Comparison with \citet{weietal13-1}}

\citet{weietal13-1} claim  $\sim$500 WDMS  binaries in their  list are
not  included in  the catalogue  by \citet{rebassa-mansergasetal12-1},
suggesting  that  our algorithm  has  failed  in  identifying a  large
fraction of  SDSS WDMS  binaries.  Here we  investigate this  issue in
detail.

The first point  to note is that out of  the $\sim$500 additional WDMS
binaries identified  by \citet{weietal13-1},  only 292 spectra  of 281
individual  objects have  been made  publicly available  following the
publication of  their paper. The  remaining $\sim$200 systems  are not
longer considered WDMS binaries by  the authors and have not been made
available by  them (Peng Wei  \& Ali Luo, private  communication).  We
visually inspect the SDSS spectra and SDSS images of these 281 systems
and find that only 85 of them are genuine WDMS binaries (see Table\,A3
for an updated classification of  the remaining 196 objects), of which
40 are new identifications  from spectroscopic plates observed as part
of  SDSS  DR\,8  (and  therefore   were  of  course  not  included  in
\citealt{rebassa-mansergasetal12-1}).  The  45 remaining WDMS binaries
were obtained  on spectroscopic  plates that formally  pre-date DR\,8.
In what  follows we study  these 45 DR\,7  and 40 DR\,8  WDMS binaries
separately  and compare them  to the  list of  DR\,7 WDMS  binaries by
\citet{rebassa-mansergasetal12-1}  and the  updated DR\,8  WDMS binary
catalogue presented in this work, respectively.

Comparing the  details of  the 45 DR\,7  WDMS binaries with  our DR\,7
WDMS  binary catalogue,  we  find  only three  systems  that were  not
included  in our  list: SDSSJ083056.71+122546.6  (noisy spectrum  of a
WDMS binary), SDSSJ170014.24+242127.3 (a  WDMS binary containing a hot
white dwarf  and an  early-type M dwarf),  and SDSSJ140127.24+484841.8
(noisy spectrum  of a  WDMS binary).  The  remaining 42  WDMS binaries
that \citet{weietal13-1}  claim to be new identifications  are in fact
contained  in  our  catalogue. However,  \citet{weietal13-1}  analysed
additional SDSS  spectra (45, as two  of the 42  objects have multiple
SDSS spectroscopy) that  were not part of the  original DR\,7 release,
and  hence were  not  available to  us  at the  time  of our  analysis
\citep{rebassa-mansergasetal12-1}.    Those   45   spectra   are   not
accessible via  DR\,7 web tools and  the DR\,7 casjobs  data base, and
appear only within  DR\,8.  The modified Julian date,  plate and fibre
identifiers  of   these  45  WDMS  binary  spectra   are  provided  in
Table\,\ref{t-plt}.   The  missed three  objects  and  45 WDMS  binary
spectra are added to our catalogue.  The total number of spectroscopic
SDSS WDMS binaries therefore increases to 2316 (Table\,\ref{t-num2})

We   then  compare   the  list   of  40   DR\,8  WDMS   binaries  from
\citet{weietal13-1} with the list of new DR\,8 WDMS binaries presented
in this work and find that  all their listed objects (and spectra) are
included in our catalogue.

\section{Characterization of the SDSS DR\,8 WDMS binary sample}

As mentioned  above, the spectroscopy within DR8  focused on observing
main  sequence  and red  giant  stars,  following  a different  target
selection  compared to  the earlier  data releases.   It  is therefore
expected  that SDSS  DR\,8 WDMS  binaries are  drawn from  a different
parent    population    than   those    in    our   DR\,7    catalogue
\citep{rebassa-mansergasetal12-1}. To investigate this we obtained the
stellar     parameters    of    the     new    47     DR\,8    systems
(Table\,\ref{t-num2})\footnote{For completeness, we also provide GALEX
  ultraviolet  and   UKIDSS  infrared  magnitudes,   and  measure  the
  secondary star  radial velocities for the new  systems following the
  method   described   in   \citet{rebassa-mansergasetal08-1}.    This
  information has been added to our on-line data base of spectroscopic
  SDSS       WDMS      binaries,      publicly       available      at
  \emph{http://www.sdss-wdms.org}}.

\begin{figure}
\centering
\includegraphics[width=\columnwidth]{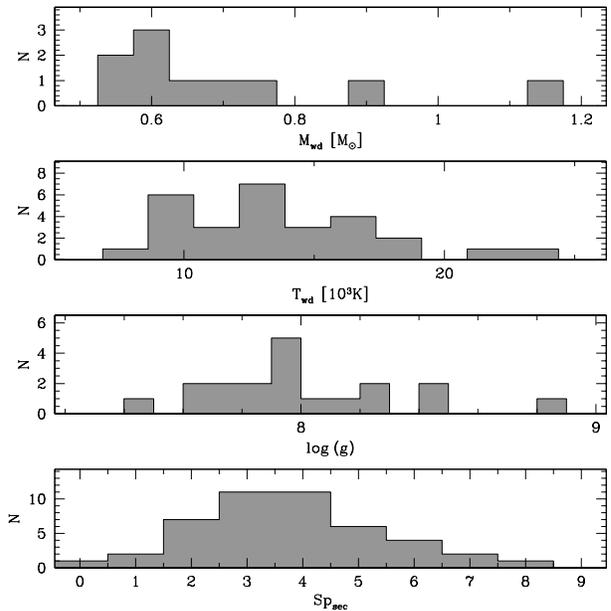}
\caption {Distributions of  white dwarf effective temperature, surface
  gravity and mass, and secondary star spectral type.}
\label{f-stellar}
\end{figure}

We obtain  the stellar parameters  following the decomposition/fitting
routine described by \citet{rebassa-mansergasetal07-1}. First, a given
SDSS WDMS binary spectrum is fitted with a two-component model using a
set of observed M dwarf  and white dwarf templates. From the converged
fit to  each WDMS binary spectrum  we record the spectral  type of the
secondary  star.   The  best-fit  M  dwarf  template,  scaled  by  the
appropriate flux  scaling factor,  is then subtracted  and we  fit the
residual white  dwarf spectrum  with a model  grid of DA  white dwarfs
\citep{koester10-1}  to obtain the  effective temperature  and surface
gravity.    From    a   mass-radius   relation    for   white   dwarfs
\citep{bergeronetal95-2,  fontaineetal01-1} we  finally  calculate the
mass of the white dwarf.

For the following  discussion we only consider white  dwarf masses and
gravities if the white  dwarf temperature exceeds 12\,000\,K.  This is
due  to a systematic  increase in  the surface  gravity that  has been
observed in recent white  dwarf spectroscopic studies below this value
\citep{koesteretal09-1,   tremblayetal11-1}.    In   order  to   avoid
contamination from unreliable stellar parameters, we additionally only
consider objects with  a relative error in the  white dwarf parameters
of less than 15 per cent.  This results in 28, 19 and 10 WDMS binaries
in the  distributions of  white dwarf effective  temperatures, surface
gravities  and  masses  respectively,  shown  on  the  top  panels  of
Figure\,\ref{f-stellar}.  The spectral types of the secondary stars of
45  of our  new  systems  are directly  determined  from the  spectral
template fitting and the corresponding distribution is provided on the
bottom panel of Figure\,\ref{f-stellar}.

\begin{figure}
\centering
\includegraphics[angle=-90,width=\columnwidth]{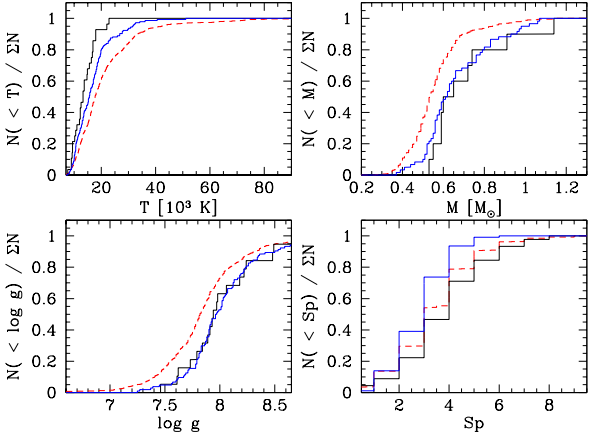}
\caption  {White dwarf  effective  temperature (top  left), mass  (top
  right), surface  gravity (bottom  left) and secondary  star spectral
  type  (bottom right)  cumulative  distributions of  SDSS DR\,8  WDMS
  binaries (black  solid lines) and  of SDSS WDMS binaries  from DR\,7
  (red  dotted   lines,  excluding  the  SEGUE   WDMS  binary  sample,
  represented as blue solid lines).}
\label{f-ks}
\end{figure}

We compare  the stellar parameter  distribution of the new  DR\,8 WDMS
binary sample  with that of the DR\,7  sample using Kolmogorov-Smirnov
(KS)  tests to the  cumulative distributions  in white  dwarf effective
temperature,  surface gravity and  mass, and  we compare  the spectral
type  distributions  using  a  $\chi^2$  test.  In  this  exercise  we
separate the  DR\,7 catalogue into  systems identified as part  of our
SEGUE  survey for  identifying  systems containing  cool white  dwarfs
(henceforth    SEGUE   WDMS    binaries,   \citealt{schreiberetal07-1,
  rebassa-mansergasetal12-1}), and stars observed  as part of the main
SDSS program (henceforth Legacy WDMS binaries).

The comparison between the DR\,8  and the Legacy WDMS binaries results
in white  dwarf parameter KS probabilities  of $\la$1 per  cent, and a
secondary star spectral type $\chi^2$ probability of 99 per cent.  The
low  KS probabilities  found  from the  comparison  of the  cumulative
distributions  of the white  dwarf parameters  are not  surprising and
confirm that  DR\,8 WDMS  binaries are drawn  from a  different parent
population: whilst  the distribution of secondary  star spectral types
is broadly  similar between  the DR8 and  Legacy WDMS  binary samples,
systematically cooler and  more massive white dwarfs are  found in the
DR\,8  sample  (see  Figure\,\ref{f-ks}).   By definition,  our  SEGUE
survey  targeted  also  WDMS  binaries containing  cool  white  dwarfs
(and/or   early-type  secondaries,  Section\,\ref{s-intro}).    It  is
therefore expected the SEGUE and DR\,8 samples to be formed by similar
white dwarf primary populations.  This hypothesis is in agreement with
the KS  probabilities of white  dwarf effective temperature,  mass and
surface gravity of  5, 76 and 96 per  cent respectively (the secondary
star spectral type  $\chi^2$ probability is of 1  per cent). Note that
the relatively  low KS  probability found from  the comparison  of the
cumulative distributions of white  dwarf effective temperatures is due
to  the   white  dwarf  components   of  DR\,8  WDMS   binaries  being
systematically cooler than  those in the SEGUE sample  (top left panel
of  Figure\,\ref{f-ks}).   Given that  WDMS  binaries containing  cool
white dwarfs are under-represented in  the current version of the SDSS
WDMS binary  catalogue, the 47 new discoveries  within DR\,8 represent
an important addition to the spectroscopic sample.

\section{Summary}
\label{s-concl}

The  current spectroscopic sample  of SDSS  WDMS binaries  is strongly
biased against systems containing  cool white dwarfs and/or early type
M dwarf companions. In this work we have provided a photometric sample
of  3\,419 SDSS  WDMS  binary  candidates that  has  the potential  of
filling in this missing and important population.  The success rate of
detecting  genuine  WDMS binaries  within  our photometric  candidates
depends on the location in colour  space and varies from 43 to 100 per
cent, with an  overall average of 84 per  cent.  The main contaminants
are cataclysmic variables and  quasars.  We estimate that the majority
($\sim$70 per  cent) of  the selected WDMS  binaries to  contain white
dwarfs   with  effective   temperatures   $\la$10\,000-15\,000\,K  and
secondary stars of spectral type $\sim$M2--3.

We have  also presented an  updated version of the  spectroscopic SDSS
WDMS binary  catalogue, which contains 2316 objects  from DR\,8.  This
is   currently   the   largest   and  most   homogeneous   sample   of
spectroscopically  identified compact  binaries.  We  identify  only a
relatively  small number,  47, of  new  WDMS binaries  within the  DR8
spectroscopy. This is due to the bulk of DR\,8 spectra being dominated
by spectra of main sequence and giant stars.  The sample of DR\,8 WDMS
binaries is clearly dominated  by systems containing cool white dwarfs
and therefore  represents an  important addition to  the spectroscopic
sample.   Stellar  parameters, magnitudes  and  secondary star  radial
velocities  of the  47  new  systems are  obtained  and made  publicly
available     via     our     SDSS     WDMS    binary     web     site
\emph{http://www.sdss-wdms.org}.

\section*{Acknowledgments}

ARM acknowledges financial support from LAMOST fellowship and Fondecyt
(3110049).  MRS acknowledges support from Milenium Science Initiative,
Chilean   Ministry  of  Economy,   Nucleus  P10-022-F,   and  Fondecyt
(1100782). The research leading  to these results has received funding
from the European Research  Council under the European Union's Seventh
Framework  Programme (FP/2007-2013)  / ERC  Grant Agreement  n. 267697
(WDTracer).   BTG  was supported  in  part  by  the UK’s  Science  and
Technology Facilities Council (ST/I001719/1).


\appendix
\section{Tables}

\begin{table*}
\centering
\caption{28  systems  in  the  list  by  \citet{liuetal12-1}  are  not
  considered WDMS binaries due to morphological problems in their SDSS
  images and/or by  visual inspection of the SDSS  spectra. We provide
  here    a    revised     classification    for    these    systems.}
\setlength{\tabcolsep}{0.8ex}
\begin{small}
\begin{tabular}{cccc}
\hline
\hline
 Object        &    class. & Object     & class.\\
\hline
SDSSJ001324.33-085021.4   &  unknown       & SDSSJ113722.24+014858.5   &  CV          \\ 
SDSSJ073721.36+464136.9   &  MS+MS sup.   & SDSSJ114955.70+284507.3   &  CV          \\ 
SDSSJ073914.18+331611.5   &  MS+MS sup.   & SDSSJ122740.82+513925.0   &  CV          \\ 
SDSSJ082619.17+110852.7   &  WD           & SDSSJ131119.61-010420.5   &  MS+MS sup.  \\ 
SDSSJ083751.00+383012.5   &  unknown       & SDSSJ132125.64+051236.4   &  M star      \\ 
SDSSJ083814.59+484134.9   &  MS+MS sup.   & SDSSJ132151.51+420014.1   &  M star+gal. \\ 
SDSSJ084008.35+490337.9   &  unknown       & SDSSJ141921.58+020710.2   &  MS+MS sup.  \\ 
SDSSJ091242.18+620940.1   &  CV           & SDSSJ145314.93+015121.9   &  M star      \\ 
SDSSJ092219.55+421256.8   &  CV           & SDSSJ154311.83+101722.2   &  MS+MS sup.  \\ 
SDSSJ094636.60+444644.7   &  CV           & SDSSJ154932.60+003524.3   &  MS+MS sup.  \\ 
SDSSJ100614.12+101620.7   &  MS+MS sup.   & SDSSJ155707.60+031813.8   &  MS+MS sup.  \\ 
SDSSJ103934.07+602744.9   &  WD           & SDSSJ162520.29+120308.8   &  CV          \\ 
SDSSJ105547.89+034555.1   &  MS+MS sup.   & SDSSJ222156.96-005118.1   &  MS+MS sup.  \\ 
SDSSJ111544.56+425822.4   &  CV           & SDSSJ235356.45-110554.2   &  MS+MS sup.  \\ 
\hline
\end{tabular}
\end{small}
\end{table*}

\begin{table*}
\centering
\caption{80  systems in  the  list by  \citet{morganetal12-1} are  not
  considered WDMS binaries due to morphological problems in their SDSS
  images and/or by  visual inspection of the SDSS  spectra. We provide
  here    a    revised     classification    for    these    systems.}
\setlength{\tabcolsep}{0.8ex}
\begin{small}
\begin{tabular}{cccccc}
\hline
\hline
 Object     & class. & Object        & class. & Object        & class.\\
\hline
SDSSJ001831.02-093139.1  & MS+MS sup.        &      SDSSJ093859.24+020925.2  & QSO+M star              &     SDSSJ152930.91+032754.0  & M star+galaxy    \\            
SDSSJ004517.26+150949.1  & MS+MS sup.        &      SDSSJ094716.59+675402.6  & morph. problems         &     SDSSJ153041.18-012008.2  & MS+MS sup.       \\            
SDSSJ005827.25+005642.6  & MS+MS sup.        &      SDSSJ100844.74+120710.3  & MS+MS sup.              &     SDSSJ160349.69+084935.8  & MS+MS sup.       \\            
SDSSJ010338.92+142538.6  & MS+MS sup.        &      SDSSJ102252.04+275828.9  & MS+MS sup.              &     SDSSJ161338.34+112740.1  & MS+MS sup.       \\            
SDSSJ012839.69-004223.4  & MS+MS sup.        &      SDSSJ103224.99+542915.4  & MS+MS sup.              &     SDSSJ161352.75+363356.1  & MS+MS sup.       \\            
SDSSJ014349.22+002130.1  & QSO+M star        &      SDSSJ110213.46+553939.4  & noisy                   &     SDSSJ161631.18+050936.2  & MS+MS sup.       \\            
SDSSJ020538.10+005835.3  & MS+MS sup.        &      SDSSJ111358.28+203206.1  & MS+MS sup.              &     SDSSJ162520.30+120308.8  & CV               \\            
SDSSJ031209.19+004701.6  & MS+MS sup.        &      SDSSJ111544.56+425822.4  & CV                      &     SDSSJ162517.58+140134.6  & MS+MS sup.       \\            
SDSSJ032131.02+000617.3  & MS+MS sup.        &      SDSSJ114036.04+575743.4  & MS+MS sup.              &     SDSSJ162702.51+252235.9  & MS+MS sup.       \\            
SDSSJ033131.33+005149.1  & MS+MS sup.        &      SDSSJ114524.45-020938.2  & MS+MS sup.              &     SDSSJ165932.19+422708.4  & MS+MS sup.       \\            
SDSSJ033436.72+005853.7  & MS+MS sup.        &      SDSSJ114653.68+012518.2  & MS+MS sup.              &     SDSSJ170112.29+193819.1  & MS+MS sup.       \\            
SDSSJ053135.46-002713.3  & MS+MS sup.        &      SDSSJ125341.55+555322.4  & MS+MS sup.              &     SDSSJ170357.74+401500.5  & MS+MS sup.       \\            
SDSSJ053233.12-001148.5  & MS+MS sup.        &      SDSSJ125324.57+555457.4  & MS+MS sup.              &     SDSSJ185808.92+192742.2  & MS+MS sup.       \\            
SDSSJ053206.23-001220.6  & MS+MS sup.        &      SDSSJ130942.34+383054.0  & MS+MS sup.              &     SDSSJ204001.99-003132.9  & MS+MS sup.       \\            
SDSSJ064941.01+290132.6  & MS+MS sup.        &      SDSSJ131227.94+161505.3  & field WD + field M star &     SDSSJ204720.07-003221.8  & MS+MS sup.       \\            
SDSSJ065012.35+273950.0  & MS+MS sup.        &      SDSSJ131954.58-011208.3  & MS+MS sup.              &     SDSSJ205732.78+010617.3  & MS+MS sup.       \\            
SDSSJ072528.10+384011.1  & MS+MS sup.        &      SDSSJ133645.56-002231.3  & MS+MS sup.              &     SDSSJ205730.72-005346.3  & MS+MS sup.       \\            
SDSSJ073531.86+315015.3  & MS+MS sup.        &      SDSSJ134554.22+005221.0  & MS+MS sup.              &     SDSSJ205857.10-075440.6  & no spectrum      \\            
SDSSJ080329.47+361934.5  & WD+red source     &      SDSSJ140236.66+542145.5  & morph. problems         &     SDSSJ210616.52-062743.3  & MS+MS sup.       \\            
SDSSJ080657.22+544546.5  & no spectrum       &      SDSSJ141325.11+515017.0  & MS+MS sup.              &     SDSSJ221410.63-002756.3  & MS+MS sup.       \\            
SDSSJ081228.68+323533.1  & unknown            &      SDSSJ143654.58-010515.4  & MS+MS sup.              &     SDSSJ222130.45+001801.8  & QSO+M star       \\            
SDSSJ081600.20+431339.2  & no spectrum       &      SDSSJ143717.41+385626.8  & morph. problems         &     SDSSJ222944.28+011323.4  & MS+MS sup.       \\            
SDSSJ083437.99+443349.4  & MS+MS sup.        &      SDSSJ144821.54+433516.9  & MS+MS sup.              &     SDSSJ223223.76+135434.5  & QSO+M star       \\            
SDSSJ083552.05+143031.7  & MS+MS sup.        &      SDSSJ145814.81+001242.7  & noisy                   &     SDSSJ233227.56+531041.5  & MS+MS sup.       \\            
SDSSJ083711.50+093828.9  & morph. problems   &      SDSSJ151923.51+073403.7  & MS+MS sup.              &     SDSSJ223530.61-000536.0  & morph. problems  \\            
SDSSJ084225.23+174453.8  & MS+MS sup.        &      SDSSJ152244.22+072945.4  & MS+MS sup.              &     SDSSJ223520.74-000558.1  & no spectrum      \\            
SDSSJ084344.93+262000.7  & MS+MS sup.        &                               &                         &     SDSSJ234456.89+010757.8  & MS+MS sup.       \\            
\hline
\end{tabular}
\end{small}
\end{table*}

\begin{table*}
\centering
\caption{196  systems  in the  list  by  \citet{weietal13-1} are  not
  considered WDMS binaries due to morphological problems in their SDSS
  images and/or by  visual inspection of the SDSS  spectra. We provide
  here    a    revised     classification    for    these    systems.}
\setlength{\tabcolsep}{0.8ex}
\begin{small}
\begin{tabular}{cccccccc}
\hline
\hline
 Object   & class. & Object    & class.  & Object    & class. \\
\hline
SDSSJ000421.61+004341.6 & noisy M star            &  SDSSJ082555.66+433526.4 & MS+MS sup.          & SDSSJ143920.72+445014.5 & unknown                 \\
SDSSJ001258.85+005920.7 & MS+MS sup.              &  SDSSJ083055.38+062554.0 & MS+MS sup.          & SDSSJ144455.91+503622.4 & MS+MS sup.             \\
SDSSJ001322.94+151457.7 & morph. problems         &  SDSSJ083202.01+124223.4 & MS+MS sup.          & SDSSJ145758.26-005549.5 & morph. problems        \\         
SDSSJ001324.33-085021.4 & MS+MS sup.              &  SDSSJ083711.50+093828.9 & morph. problems     & SDSSJ150916.11+094227.0 & MS+MS sup.             \\         
SDSSJ001601.40+000832.4 & MS+MS sup.              &  SDSSJ083751.00+383012.5 & unknown              & SDSSJ151923.51+073403.7 & MS+MS sup.             \\         
SDSSJ001610.17-002421.6 & MS+MS sup.              &  SDSSJ084224.78+023907.1 & MS+MS sup.          & SDSSJ152244.22+072945.4 & MS+MS sup.             \\         
SDSSJ001658.16-101108.4 & MS+MS sup.              &  SDSSJ084225.23+174453.8 & MS+MS sup.          & SDSSJ152458.89+585131.5 & morph. problems        \\         
SDSSJ002334.83+004021.9 & MS+MS sup.              &  SDSSJ084324.98+104110.7 & MS+MS sup.          & SDSSJ153119.09-023008.0 & MS+MS sup.             \\         
SDSSJ004517.26+150949.1 & morph. problems         &  SDSSJ084344.93+262000.7 & MS+MS sup.          & SDSSJ153238.62+003435.2 & MS+MS sup.             \\         
SDSSJ005444.02-002120.1 & MS+MS sup.              &  SDSSJ085057.18+381941.7 & MS+MS sup.          & SDSSJ153243.35+312057.3 & morph. problems        \\         
SDSSJ005827.25+005642.6 & MS+MS sup.              &  SDSSJ090155.27+123710.7 & MS+MS sup.          & SDSSJ153350.13-011007.2 & MS+MS sup.             \\         
SDSSJ011248.52+001132.2 & unknown                  &  SDSSJ090323.57+470406.9 & morph. problems     & SDSSJ153434.40+023801.2 & MS+MS sup.             \\         
SDSSJ012127.23-092905.0 & M star + back.          &  SDSSJ090848.27+335354.8 & morph. problems     & SDSSJ153453.47+263238.7 & morph. problems        \\         
SDSSJ012839.69-004223.4 & MS+MS sup.              &  SDSSJ091242.18+620940.1 & CV                  & SDSSJ154544.60+094823.9 & MS+MS sup.             \\         
SDSSJ013701.06-091234.9 & CV                      &  SDSSJ093919.09+274314.0 & WD                  & SDSSJ154653.68+573533.8 & MS+MS sup.             \\         
SDSSJ015151.87+140047.2 & CV                      &  SDSSJ094056.07+095408.6 & morph. problems     & SDSSJ155156.59+352928.1 & MS+MS sup.             \\         
SDSSJ020357.55-005424.6 & MS+MS sup.              &  SDSSJ094636.60+444644.8 & CV                  & SDSSJ155349.21+394106.2 & unknown                 \\         
SDSSJ020605.67-001723.7 & MS+MS sup.              &  SDSSJ095228.57+540340.0 & morph. problems     & SDSSJ155412.34+272152.4 & CV                     \\         
SDSSJ021041.54-001739.2 & noisy M star            &  SDSSJ095306.14+250905.5 & MS+MS sup.          & SDSSJ155644.24-000950.2 & CV                     \\         
SDSSJ025448.73+002310.6 & MS+MS sup.              &  SDSSJ100614.12+101620.7 & MS+MS sup.          & SDSSJ160349.69+084935.8 & MS+MS sup.             \\         
SDSSJ025712.15+002537.7 & morph. problems         &  SDSSJ100844.74+120710.3 & MS+MS sup.          & SDSSJ160420.96+512734.5 & MS+MS sup.             \\         
SDSSJ025750.13-002757.3 & MS+MS sup.              &  SDSSJ100904.57+114630.8 & M star + blue excess& SDSSJ160839.59+450358.5 & morph. problems        \\         
SDSSJ030534.15+385308.0 & morph. problems         &  SDSSJ102329.04+092926.1 & broken spectrum     & SDSSJ160941.64+525332.7 & MS+MS sup.             \\         
SDSSJ030739.91+003656.0 & MS+MS sup.              &  SDSSJ103110.75+425709.9 & MS+MS sup.          & SDSSJ162517.58+140134.6 & morph. problems        \\         
SDSSJ031104.85+412109.4 & MS+MS sup.              &  SDSSJ103301.59+091757.2 & morph. problems     & SDSSJ162520.30+120308.8 & CV                     \\         
SDSSJ033131.33+005149.1 & MS+MS sup.              &  SDSSJ103623.24+081007.0 & morph. problems     & SDSSJ162702.51+252235.9 & MS+MS sup.             \\         
SDSSJ034538.00-064326.2 & M star + galaxy         &  SDSSJ103738.28-002328.9 & MS+MS sup.          & SDSSJ163037.11+165855.8 & MS+MS sup.             \\         
SDSSJ044403.97-054659.9 & MS+MS sup.              &  SDSSJ104730.07+605457.4 & WD                  & SDSSJ163809.60+453308.6 & noisy M star           \\         
SDSSJ045325.66-054459.1 & MS+MS sup.              &  SDSSJ105018.43+423406.3 & MS+MS sup.          & SDSSJ165343.40+630549.3 & morph. problems        \\         
SDSSJ053125.11-001850.8 & morph. problems         &  SDSSJ105420.35+163154.6 & MS+MS sup.          & SDSSJ170053.30+400357.6 & unknown                 \\         
SDSSJ053135.46-002713.3 & noisy M star            &  SDSSJ105707.25+261416.7 & active M star       & SDSSJ170112.29+193819.1 & MS+MS sup.             \\         
SDSSJ053201.53-002916.1 & M star + blue excess    &  SDSSJ110330.10+323236.0 & M star + galaxy     & SDSSJ170213.26+322954.1 & CV                     \\         
SDSSJ053206.23-001220.6 & morph. problems         &  SDSSJ110555.86-165634.4 & MS+MS sup.          & SDSSJ170259.78+201609.2 & MS+MS sup.             \\         
SDSSJ053233.12-001148.5 & M star + blue excess    &  SDSSJ111358.28+203206.1 & morph. problems     & SDSSJ170919.90+612016.8 & morph. problems        \\         
SDSSJ053515.70-011050.9 & M star                  &  SDSSJ111544.56+425822.4 & CV                  & SDSSJ172943.51+330220.2 & MS+MS sup.             \\         
SDSSJ053528.67-011414.7 & M star + blue excess    &  SDSSJ113722.25+014858.6 & CV                  & SDSSJ174014.75+551157.9 & MS+MS sup.             \\         
SDSSJ053530.27-010720.6 & M star + blue excess    &  SDSSJ114036.04+575743.5 & morph. problems     & SDSSJ191404.82+781725.3 & MS+MS sup.             \\         
SDSSJ053617.85-005507.1 & M star                  &  SDSSJ120544.40+614103.9 & MS+MS sup.          & SDSSJ191956.88+382900.4 & MS+MS sup.             \\         
SDSSJ053619.98-010107.0 & M star + blue excess    &  SDSSJ120756.53+032356.3 & MS+MS sup.          & SDSSJ192840.06+792244.7 & MS+MS sup.             \\         
SDSSJ053639.15-005905.2 & M star                  &  SDSSJ121226.69+252158.4 & MS+MS sup.          & SDSSJ193822.75+621212.3 & noisy M star           \\         
SDSSJ053806.74-010817.8 & active M star           &  SDSSJ122454.76+125212.4 & morph. problems     & SDSSJ200942.51-125234.9 & MS+MS sup.             \\         
SDSSJ060910.64+241858.5 & morph. problems         &  SDSSJ123032.23+042838.2 & MS+MS sup.          & SDSSJ201808.19+754654.7 & MS+MS sup.             \\         
SDSSJ065056.13+164319.9 & morph. problems         &  SDSSJ123932.01+210806.3 & unknown              & SDSSJ203708.43+151909.1 & MS+MS sup.             \\         
SDSSJ070105.84+292941.9 & MS+MS sup.              &  SDSSJ124240.71+135540.7 & early type star     & SDSSJ204322.39-003104.4 & noisy M star           \\         
SDSSJ072528.10+384011.1 & unknown                  &  SDSSJ124548.48+231346.7 & morph. problems     & SDSSJ210616.52-062743.3 & MS+MS sup.             \\         
SDSSJ072918.08+414138.0 & MS+MS sup.              &  SDSSJ125324.57+555457.4 & morph. problems     & SDSSJ211748.66+115406.9 & MS+MS sup.             \\         
SDSSJ073430.01+314013.7 & morph. problems         &  SDSSJ125341.55+555322.4 & morph. problems     & SDSSJ213327.56-000929.5 & MS+MS sup.             \\         
SDSSJ073442.76+314449.3 & morph. problems         &  SDSSJ125526.05+381438.0 & MS+MS sup.          & SDSSJ214854.81-070300.9 & MS+MS sup.             \\         
SDSSJ073447.51+314510.0 & morph. problems         &  SDSSJ125526.45+381656.5 & morph. problems     & SDSSJ215457.67-000120.0 & MS+MS sup.             \\         
SDSSJ073531.86+315015.3 & MS+MS sup.              &  SDSSJ125918.24+235632.7 & MS+MS sup.          & SDSSJ215503.22-010045.3 & noisy M star           \\         
SDSSJ073721.35+464136.9 & MS+MS sup.              &  SDSSJ130942.34+383054.0 & morph. problems     & SDSSJ220447.19-005324.8 & morph. problems        \\         
SDSSJ074505.56+315054.9 & MS+MS sup.              &  SDSSJ131954.58-011208.3 & MS+MS sup.          & SDSSJ221410.59-002754.5 & MS+MS sup.             \\         
SDSSJ075117.10+144423.6 & M star + blue excess    &  SDSSJ132517.49+550020.3 & morph. problems     & SDSSJ221949.37+005731.2 & WD                     \\         
SDSSJ075227.56+182957.5 & MS+MS sup.              &  SDSSJ132935.64-022711.9 & MS+MS sup.          & SDSSJ223520.74-000558.1 & morph. problems        \\         
SDSSJ075339.62+353213.2 & unknown                  &  SDSSJ133841.30+101602.2 & unknown              & SDSSJ223530.61-000536.0 & morph. problems        \\         
SDSSJ075546.51+394754.6 & noisy M star            &  SDSSJ134159.10+041224.0 & MS+MS sup.          & SDSSJ225703.39-010457.8 & MS+MS sup.             \\         
SDSSJ075605.79+664505.5 & MS+MS sup.              &  SDSSJ134748.79+235153.0 & MS+MS sup.          & SDSSJ230949.12+213516.7 & active M star          \\         
SDSSJ075633.87+355534.2 & MS+MS sup.              &  SDSSJ135554.18+070822.5 & MS+MS sup.          & SDSSJ231808.91-091145.3 & morph. problems        \\         
SDSSJ075732.50+243024.2 & MS+MS sup.              &  SDSSJ140405.76+641507.6 & unknown              & SDSSJ232237.76+135859.2 & MS+MS sup.             \\         
SDSSJ080045.22+402318.5 & WD                      &  SDSSJ140550.20+231113.4 & unknown              & SDSSJ232753.44+133855.5 & MS+MS sup.             \\         
SDSSJ080104.16+391320.9 & MS+MS sup.              &  SDSSJ140550.31+641947.0 & unknown              & SDSSJ233010.65+535854.4 & MS+MS sup.             \\         
SDSSJ080140.00+250617.0 & morph. problems         &  SDSSJ141325.11+515017.1 & morph. problems     & SDSSJ233227.37+531051.5 & morph. problems        \\         
SDSSJ080237.16+134909.5 & noisy M star            &  SDSSJ141710.11+103415.8 & M star + galaxy     & SDSSJ234339.71-004852.5 & MS+MS sup.             \\         
SDSSJ080329.47+361934.5 & morph. problems         &  SDSSJ142830.48+013443.6 & MS+MS sup.          & SDSSJ234811.79+140926.4 & G star                 \\         
SDSSJ081245.66+293840.7 & MS+MS sup.              &  SDSSJ143332.31+414117.0 & MS+MS sup.          &                         &                        \\   
SDSSJ081751.00+324340.5 & unknown                  &  SDSSJ143654.58-010515.4 & MS+MS sup.          &                         &                        \\
\hline
\end{tabular}
\end{small}
\end{table*}

\end{document}